\documentclass[12pt]{iopart}

\usepackage{iopams}
\usepackage[pdftex]{graphicx}
\usepackage[section] {placeins}
\usepackage{multirow}
\usepackage{float}
\usepackage[innercaption]{sidecap}
\usepackage{amsfonts}
\usepackage{amssymb}
\usepackage{bm}
\begin{document}

\newcommand{\FeNi}{Ba(Fe$_{1-x}$Ni$_{x}$)$_2$As$_2$ }
\newcommand{\FeCo}{Ba(Fe$_{1-x}$Co$_{x}$)$_2$As$_2$ }
\newcommand{\FeT}{Ba(Fe$_{1-x}$T$_{x}$)$_2$As$_2$ }
\newcommand{\BaFe}{BaFe$_2$As$_2$}

\title[London Penetration Depth in Pnictide Superconductors ]{London Penetration Depth in Iron - based Superconductors}

\author{Ruslan Prozorov}

\address{The Ames Laboratory\\
Department of Physics \& Astronomy, Iowa State University\\
Ames, IA 50011, U.S.A.\\
prozorov@ameslab.gov}

\author{Vladimir~G.~Kogan}

\address{The Ames Laboratory\\
Ames, IA 50011, U.S.A.\\
kogan@ameslab.gov}

\date{4 July 2011}

\begin{abstract}
Measurements of London penetration depth are a sensitive tool to study multi-band superconductivity and it has provided several important insights to the behavior of Fe-based superconductors. We first briefly review the ``experimentalist - friendly'' self-consistent  Eilenberger model that relates the measurable superfluid density and structure of superconducting gaps. Then we focus on the \BaFe~-derived materials, for which the results are consistent with 1) two distinct superconducting gaps; 2) development of strong in-plane gap anisotropy with the departure from the optimal doping; 3) appearance of gap nodes along the $c-$direction in a highly overdoped regime; 4) significant pair-breaking, presumably due to charge doping; 5) fully gapped (exponential) intrinsic behavior at the optimal doping if scattering is removed (probed in the ``self-doped'' stoichiometric LiFeAs); 6) competition between magnetically ordered state and superconductivity, which do coexist in the underdoped compounds. Overall, it appears that while there are common trends in the behavior of Fe-based superconductors, the gap structure is non-universal and is sensitive to the doping level. It is plausible that the rich variety of possible gap structures within the general $s_{\pm}$ framework is responsible for the observed behavior.
\end{abstract}


\section{Introduction}
\paragraph{}
After the initial discoveries of superconductivity in Fe-based superconductors, first in LaFePO with $T_c\approx4$ K \cite{Kamihara2006LaFePO} and then in LaFeAs(O$_{1-x}$F$_x$) with $T_c \approx 26$ K \cite{Kamihara2008LaFeAsFO}, $T_c$ as high as 55 K has been reported in SmFeAs(O$_{1-x}$F$_x$) \cite{Ren2008Sm1111}. These findings have triggered an intense research aimed at understanding the fundamental physics governing this new family of superconductors. Since the original discoveries,  at least seven different classes of iron-based superconductors  have been identified, all exhibiting unconventional physical properties \cite{Johnston2010review,Mazin2010Nature,Paglione2010review,Stewart2011RMP}. Among those, perhaps the most diverse family is based on (AE)Fe$_2$As$_2$ parent compounds (where AE=alkaline earth, denoted the ``122" system) in which electron \cite{Sefat2008FeCo122,Ni2010FeT122,Ni2010PdRh122,Canfield2010review122},
hole \cite{Rotter2008BaK122,Ni2008BaK122} and isovalent \cite{Ren2009EuAsP122,Jiang2009BaAsP122} dopants induce a superconductivity ``dome'' with maximum $T_c$ achieved at fairly high concentrations of the dopant (up to 45 \% in case of BaK122) and showing coexistence of superconductivity and long-range magnetic order on the underdoped side.

Three plus years into the intense study of the Fe-based superconductors, several experimental conclusions supported by a large number of reports can be made. Among those, relevant to our discussion of London penetration depth, are: 1) two distinct superconducting gaps of the magnitude ratio of about 2; 2) power-law behavior of thermodynamic quantities at low temperatures presumably due to pair-breaking scattering and gap anisotropy; 3) doping  dependent three-dimensional gap structure possibly with nodes. For some reviews of various experiments and theories the reader is referred to
\cite{Johnston2010review,Mazin2010Nature,Paglione2010review,Stewart2011RMP,Mizuguchi2010review11,Wilson2010JPCM,Maiti2011,Canfield2010review122,Hosono2008review}. Here we focus on London penetration depth as a sensitive tool to study superconducting gap structure, albeit averaged over the Fermi surface \cite{Prozorov2006SUSTreview}.

\section{Measurements of London Penetration Depth}

\paragraph{}There are several ways to measure magnetic penetration depth in superconductors \cite{Prozorov2006SUSTreview}.
These include muon-spin rotation ($\mu$SR) \cite{Luetkens2008La1111,Williams2010FeCouSR,Sonier2010}, frequency  dependent conductivity \cite{ValdesAguilar2010THzFeCo122,Wu2010IR}, magnetic force and SQUID microscopy \cite{Luan2010FeCo122Opt,Luan2011FeCoL0}, measurements of the first critical field using either global \cite{Prozorov2009physC,Song2010LiFeAsHc1} or local probes \cite{Okazaki2009Pr1111,Klein2010FeSeTe}, microwave cavity perturbation \cite{Hashimoto2009BaK122,Hashimoto2009Pr1111,Shibauchi2009review,Bobowski2010MW,Takahash2011MWFeSeTe},  mutual inductance (especially suitable for thin films) \cite{Yong2011FeCo122films} and self-oscillating tunnel-diode resonator (TDR) \cite{Prozorov2009physC,Malone2009Sm1111,Gordon2010FeCoL0vsX,Gordon2010Pairbreaking,Kim2011LiFeAsLambda,Martin2010FeNinodes,Martin2009NdLa1111,Shibauchi2009review}. Each technique has its advantages and disadvantages and only by a combination of several different measurements performed on different samples and at different experimental conditions, one can obtain a more or less objective picture. In the case of Fe-based superconductors, most of the reports agree in general, but still differ on the details. Most difficulties come from the necessity to resolve very small variations of the penetration depth at the lowest temperatures (below 1 K) while maintaining extreme temperature resolution and the ability to measure statistically sufficient number of data points over the full temperature range. In addition, it is important to look at a general picture by measuring many samples of each composition and over extended doping regime. This review is based on the results obtained with a self-oscillating tunnel-diode resonator (TDR) that has the advantage of providing stable resolution of about 1 part per 10$^9$ \cite{Prozorov2006SUSTreview}. Indeed, comparison of TDR with other techniques, especially bulk probes, such as thermal conductivity, is important to come up with a consensus regarding the gap structure in pnictide superconductors \cite{Luo2009BaK122TC,Tanatar2010FeCoTCvsX,Reid2010FeCo122TC}.

\subsection{Tunnel - diode resonator (TDR)}

\subsubsection{Frequency - domain measurements:}\label{SEC:TDR}

\paragraph{}When a non-magnetic conducting sample is inserted into a coil of a tank circuit of quality factor $Q$ and resonating at a frequency $f_{0}=1/2\pi\sqrt{L_0C}$, it causes changes in the resonant frequency and the quality factor. For a flat slab of width $2w$ and volume $V_s$ in parallel field \cite{Hardy1993YBCO}:

\begin{equation}
\frac{\Delta f}{f_0} = \frac{f_0-f\left( \alpha \right)}{f_0}=\frac{V_{s}}{2V_{c}}\left( 1-\Re\frac{\tanh \left(\alpha w\right) }{\alpha w}\right)
\end{equation}%

\begin{equation}
\Delta \left( \frac{1}{Q}\right) =\frac{V_{s}}{V_{c}}\Im\left(\frac{\tanh{\left( \alpha w\right)} }{\alpha w}\right)
\end{equation}%

\noindent where $\alpha =(1-i)/\delta$ in a normal metal and $\alpha =1/\lambda$ in a superconductor. Here, $\delta $ is the normal skin depth and $\lambda $ is the London penetration depth. $V_{c}$ is volume of the inductor. Solving explicitly, we have for a normal metal:

\begin{equation}
\frac{\Delta f}{f}=\frac{V_{s}}{2V_{c}}\left( 1-\frac{\delta }{2w}\frac{%
\sinh \frac{2w}{\delta }+\sin \frac{2w}{\delta }}{\cosh \frac{2w}{\delta }%
+\cos \frac{2w}{\delta }}\right)
\end{equation}

\noindent and, in a superconductor, we obtain the known ``infinite slab'' solution:

\begin{equation}
\frac{\Delta f}{f}=\frac{V_{s}}{2V_{c}}\left( 1-\frac{\lambda }{w}\tanh
\left( \frac{w}{\lambda }\right) \right)
 \end{equation}%


\noindent which allows us to measure both superconducting penetration depth and normal - state skin depth (which gives the contact-less measure of resistivity).

In the case of a finite sample with magnetic susceptibility $\chi$, we can still write: $\Delta f =-4 \pi\chi(T)G$, where $G\simeq f_{0}V_{s}/2V_{c}\left( 1-N\right) $ is a geometric calibration constant that includes the effective demagnetization factor $N$ \cite{Prozorov2000Meissner}. Constant $G$ is measured directly by pulling the sample out of the coil \cite{Prozorov2000Meissner,Prozorov2006SUSTreview}. The susceptibility in the Meissner state can be written in terms of London penetration depth $\lambda(T)$ and normal state paramagnetic permeability (if local - moment magnetic impurities are present), $\mu(T)$, as \cite{Prozorov2000Meissner}:

\begin{equation}
4\pi\chi=\frac{\sqrt{\mu(T)} \lambda(T)}{R}\tanh{\left(\frac{\sqrt{\mu(T)} R}{ \lambda(T)}\right)}-1
\label{TDR}
\end{equation}

\noindent where $R$ is the characteristic sample dimension. For a most typical in the experiment slab geometry with $2b \geq 2a$ planar dimensions and thickness $2d$ (magnetic field excitation is along the $d-$ side), $R$ is approximately given by \cite{Prozorov2000Meissner}:

\begin{equation}
\frac{w}{R}\approx 2 \left[1+\left(1+\left(\frac{2d}{w}\right)^2 \right)\arctan{\left(\frac{w}{2d}\right)}-\frac{2d}{w} \right]
\label{R}
\end{equation}

\noindent with $w \approx ab/(a+b)$.

Unlike microwave cavity perturbation technique that requires scanning the frequency \cite{Hardy1993YBCO}, TDR is a self-oscillating resonator always locked onto its resonant frequency \cite{Degrift1975TDR}. A sample to be studied is mounted on a sapphire rod and inserted into the inductor coil of the tank circuit. Throughout the measurement the temperature of the circuit (and of the coil) is stabilized at $\pm 1$ mK. This is essential for the stability in the resonant frequency, which is resolved to about 0.01 Hz. This translates to the ability to detect changes in $\lambda(T)$ in the range of a few {\AA}ngstrom. The ac magnetic excitation field, $H_{ac}$, in the coil is about 20 mOe, which
is small enough to ensure that no vortices are created and London penetration depth is measured.

\subsubsection{Measurements of the absolute value of $\lambda(T)$:}\label{SEC:L0}

\paragraph{}The described TDR technique provides precise measurements of the \emph{variation} of the penetration depth, $\Delta\lambda(T)$, but not the absolute value for the reasons described in detail elsewhere \cite{Prozorov2000L0}. However, the TDR technique can be extended to obtain the absolute value of the penetration depth, $\lambda(T)$. The idea is to coat the entire surface of the superconductor under study with a thin film of a conventional superconductor with lower $T_c$ and a known value of $\lambda(0)$. In this work we used aluminum, which is convenient, since its $T_c^{Al} \approx 1.2$ K is quite low for most of the discussed materials, so it is possible to extrapolate to $T = 0$ and obtain $\lambda(0)$. While the Al film is superconducting it screens the magnetic field and the effective penetration depth is \cite{Gordon2010FeCoL0vsX}:

\begin{equation}
\label{LeffL0}
\lambda_{eff}(T)=\lambda_{Al}(T)\frac{\lambda(T)+\lambda_{Al}(T)\tanh{\left(\frac{t}{\lambda_{Al}(T)}\right)}}
{\lambda_{Al}(T)+\lambda(T)\tanh{\left(\frac{t}{\lambda_{Al}(T)}\right)}},
\end{equation}

\noindent However, when it becomes normal, Al layer does not cause any screening because its thickness, $t$, is much less than the normal state skin depth at the TDR operating frequency of 14 MHz, where $\delta_{Al} \approx$ 75 $\mu$m for $\rho^{Al}_0$=10 $\mu \Omega$-cm \cite{Hauser1972Al}. By measuring the frequency shift upon warming from the base temperature, $T = T_{min}$, to $T = T_c^{Al}$ where $\lambda_{Al} \rightarrow \infty$ and Eq.~(\ref{LeffL0}) gives $\lambda_{eff}(T_c^{Al}) = t+\lambda(T_c^{Al})$, we can calculate $\lambda(T_c^{Al})$.  We also note that if aluminum coating is damaged (i.e., cracks and/or holes), the result will \emph{underestimate} $\lambda(0)$. This method, therefore, provides the lower boundary estimate.

\subsubsection{Out-of-plane Penetration Depth:}
\label{SEC:Lc}

\paragraph{}Finally, the TDR technique can be used to measure out-of-plane component of the penetration depth, $\lambda_c$. For the excitation field $H_{ac}\parallel c$, screening currents flow only in the $ab$-plane and $\Delta f$ is only related to the in-plane penetration depth, $\Delta\lambda_{ab}$. However, when the magnetic field is applied along the $ab$-plane ($H_{ac}\parallel ab$), screening currents flow both in the plane and between the planes, along the $c$-axis. In this case, $\Delta f^{\perp}$ contains contributions from both $\Delta\lambda_{ab}$ and $\Delta\lambda_{c}$. For a rectangular sample of thicknesses $2t$, width $2w$ and length $l$, $\Delta f^{\perp}$ is approximately given by Eq.~(\ref{eqmix})

\begin{equation}
\frac{\Delta f^{\perp}}{\Delta f^{\perp}_{0}} \approx \frac{\Delta\lambda_{ab}}{t}+\frac{\Delta\lambda_{c}}{w}=\frac{\Delta\lambda_{mix}}{R_b}
\label{eqmix}
\end{equation}

\noindent where $R_b$ is the effective dimension that takes into account finite size
effects \cite{Prozorov2000Meissner}, Eq.~\ref{R}. Knowing $\Delta\lambda_{ab}$ from the measurements with $H_{ac}\parallel c$ and the sample dimensions, one can obtain $\Delta\lambda_{c}$ from Eq.~(\ref{eqmix}). However, because $2w\geq 4\times 2t$ in most cases, $\Delta f^{\perp}$ is in general dominated by the contribution from $\Delta\lambda_{ab}$. The subtraction of $\Delta\lambda_{c}$ becomes therefore prone to large errors. The alternative, more accurate approach is to measure the sample twice \cite{Fletcher2007NbSe2}. After the first measurement with the field along the longest side $l$ ($H_{ac}\parallel l$), the sample is cut along this $l$ direction in two halves, so that the width (originally $2w$) is reduced to $w$. Since the thickness $2t$ remains the same, we can now use Eq.~(\ref{eqmix}) to calculate $\Delta\lambda_{c}$ without knowing $\Delta\lambda_{ab}$. Note that the length $l$ and width $w$ are in the crystallographic $ab-$plane, whereas the thickness $2t$ is measured along the $c-$axis. In our experiments, both approaches to estimate $\Delta \lambda_c(T)$ produced similar temperature dependence, but the former technique had a larger data scatter, as expected.

\section{London Penetration Depth and Superconducting Gap}\label{SEC:theory}

\paragraph{}In order to describe London penetration depth in multi-gap superconductors and to take into account pair-breaking scattering, we use the weak coupling Eilenberger quasi-classical formulation of the superconductivity theory that holds for a general anisotropic Fermi surface and for any gap symmetry \cite{Eilenberger1968}. This method suitable for the analysis of the experimental data is described in a few publications of one of us\cite{Kogan2002anisotropy,Kogan2009pairBreaking,Kogan2009gamma,Gordon2010Pairbreaking};  here we briefly summarize the scheme.

 In the clean case the Eilenberger equations read:

\begin{eqnarray}
{\bm v} {\bm  \Pi}f=2\Delta g/\hbar  -2\omega f \,,
\label{eil1}\\
-{\bm v} {\bm  \Pi}^*f^+=2\Delta^* g/\hbar  -2\omega f^+ \,,
\label{eil2}\\
g^2=1-ff^{+}\,, \label{eil3}\\
\Delta({\bm  r},{\bm v})=2\pi TN(0) \sum_{\omega >0}^{\omega_D} \Big\langle
V({\bm v},{\bm v}^{\prime\,}) f({\bm v}^{\prime},{\bm
r},\omega)\Big\rangle_{{\bm v}^{\prime\,}},   \label{eil4}\\
{\bm  j}=-4\pi |e|N(0)T\,\, {\rm Im}\sum_{\omega >0}\Big\langle {\bm v}g\Big\rangle\,.
\label{eil5}
\end{eqnarray}

\noindent Here, ${\bm v}$ is the Fermi velocity, ${\bm  \Pi} =\nabla +2\pi i{\bm
A}/\phi_0$, $\phi_0$ is the flux quantum. $\Delta ({\bm  r})$ is the gap
function (the order parameter) which may depend on the position ${\bm  k}_F$ at
the Fermi surface (or on ${\bm v}$).  Eilenberger functions $f({\bm  r},{\bm v},\omega),\,\,  f^{+} $, and $g$ originate from Gor'kov  Green's functions integrated over the energy variable near the Fermi
surface to exclude fast spatial oscillations on the scale $1/k_F$; instead
$f,g$ vary on the relevant for superconductivity  scale of the
coherence length $\xi$. Functions $f, f^+$ describe the superconducting
condensate, whereas $g$ represents normal excitations. $N(0)$ is the total density of states at the Fermi level per spin. The Matsubara frequencies are defined by $\hbar\omega=\pi T(2n+1)$
with an integer $n$, and $\omega_D$ being the Debye frequency (for phonon-mediated superconductivity or a relevant energy scale for other mechanisms). The averages over the Fermi surface weighted with the local density of states
$\propto 1/|{\bm v}|$ are defined as

\begin{equation}
\Big\langle X \Big\rangle = \int \frac{d^2{\bm  k}_F}{(2\pi)^3\hbar N(0)|{\bm v}|}\,
\,X\,.
\label{<>}
\end{equation}

The order parameter $\Delta$ is related to $f$ in the self-consistency
equation Eq.~(\ref{eil4}). Often, the effective coupling $V$ is assumed to be factorizable \cite{MarkowitzKadanoff1963}, $ V({\bm v},{\bm v}^{\prime\,})=V_0 \,\Omega({\bm v})\,\Omega({\bm v}^{\prime\,})$; this assumption is not always justifiable but it makes the algebra much simpler. One  looks for the order parameter in the form  $\Delta (
{\bm  r},T;{\bm v})=\Psi ({\bm  r},T)\, \Omega({\bm v})$.
Then, the self-consistency Eq.\,(\ref{eil4}) takes the form:

\begin{equation}
\Psi( {\bm  r},T)=2\pi T N(0)V_0 \sum_{\omega >0}^{\omega_D} \Big\langle
\Omega({\bm v} ) f({\bm v} ,{\bm  r},\omega)\Big\rangle \,.
\label{gap}
\end{equation}
The  function $\Omega({\bm v})$ (or $\Omega({\bm  k}_F)$), which describes the
variation of $\Delta$ along the Fermi surface, is conveniently normalized \cite{Pokrovskii1961anisoSC}:

\begin{equation}
       \Big\langle \Omega^2 \Big\rangle=1\,.
\label{norm}
\end{equation}

In the absence of currents and fields ${\bm  \Pi}=0$, and the Eilenberger
equations give for the uniform ground state:
\begin{equation}
f _0 =f ^+_0={\Delta_0 \over
\beta },\quad g _0  =  {\hbar\omega\over
\beta } ,\quad
\beta ^2=\Delta_0 ^2+ \hbar^2\omega^2\,; \label{f_0}
\end{equation}
Note that in  general, both $\Delta_0=\Psi_0(T) \Omega(\bm k_F)$ and
$\beta$ depend  on the position ${\bm  k}_F$ at the Fermi surface.

Instead of dealing with the effective microscopic electron-electron
interaction $V$, one can use in this formal scheme the measurable critical
temperature $T_c$ utilizing the identity
\begin{equation}
\frac{ 1}{N(0) V_0}= \ln \frac{T}{T_{c}}+2\pi T\sum_{\omega >0}^{\omega_D}
       \frac{ 1}{\hbar \omega} \,,
\label{1/NV}
\end{equation}
which is equivalent to the famous relation $\Delta(0)=\pi T_{c}
e^{-\gamma}=2\hbar\omega_D
\exp(-1/N(0)V_0)$; $\gamma=0.577$ is the Euler constant.  Substitute Eq.\,(\ref{1/NV}) in
Eq.\,(\ref{gap}) and replace
$\omega_D$ with infinity due to  fast convergence:
       \begin{equation}
\frac{\Psi }{2\pi T} \ln \frac{T_{c}}{T}= \sum_{\omega
>0}^{\infty}\left(\frac{\Psi}{\hbar\omega}-\Big\langle \Omega \, f
\Big\rangle
\right)\,.
\label{gap1}
\end{equation}
Now equation for $\Psi(T)$ reads:
\begin{equation}
\frac{1}{2\pi T} \ln \frac{T_{c}}{T}= \sum_{\omega
>0}^{\infty}\left(\frac{1}{\hbar\omega}-\Big\langle
\frac{\Omega^2}{\sqrt{\Psi^2\Omega^2+\hbar^2\omega^2}}\Big\rangle
\right)\,.
\label{gap2}
\end{equation}

\subsection{London Penetration Depth}

\paragraph{}
Within microscopic theory, the penetration of weak magnetic fields into
 superconductors is treated perturbatively.
Weak supercurrents and fields leave the order parameter modulus unchanged,
but  cause the condensate, i.e., $\Delta$ and  the amplitudes $f$ to acquire an
overall phase $\theta({\bm  r})$. Using the method of perturbations, one obtains
corrections to $f_0,g_0$ among which we need only
       \begin{equation}
g _1 =i\hbar\,\frac{\Delta_0 ^ 2 }{2\beta ^3}\,
       {\bm v}\cdot(\nabla\theta+ 2\pi{\bm  A}/\phi_0 )  \,.
\label{g_corr}
\end{equation}
Substituting this in the general expression (\ref{eil5}) for the current
density, one obtains
the London relation between the current and the ``gauge invariant vector potential" ${\bm  a}=  \phi_0\nabla\theta/2\pi + {\bm  A}$:
\begin{equation}
  4\pi j_i/c=-  (\lambda^2)_{ik}^{-1} a_k\,.
\end{equation}
Then, the general expression (\ref{eil5}) for the current  gives the inverse tensor of the squared penetration depth:
\begin{equation}
(\lambda^2)_{ik}^{-1}= \frac{16\pi^2 e^2T}{
c^2}\,N(0)  \sum_{\omega} \Big\langle\frac{
\Delta_0^2v_iv_k}{\beta ^{3}}\Big\rangle \,.
\label{lambda-tensor}
\end{equation}
This result holds at any temperature for clean materials with arbitrary Fermi surface and order parameter anisotropies \cite{Kogan2002anisotropy}. The temperature dependence of   $\Delta_0=\Psi\Omega$ and $\beta$ should be obtained solving
Eq.\,(\ref{gap2}).

Thus, the general scheme for evaluation of $\lambda(T)$ in clean superconductors consists of two major steps: first evaluate the order
parameter $\Delta_0(T)$ in uniform zero-field state for a given gap anisotropy $\Omega({\bm v})$, then use  Eq.\,(\ref{lambda-tensor}) with a proper averaging over the Fermi surface. The sum over Matsubara frequencies is rapidly convergent and is easily done numerically (except in a few limiting situations for which analytic  evaluation is possible).

\subsubsection{Isotropic $\Delta$ on a general Fermi surface: }

For majority of materials with electron-phonon interaction responsible for superconductivity, the relevant phonons have frequencies   on the order of $\omega_D$. One can see that in exchanging such phonons, the electron momentum transfer is of the order of $\hbar k_F$; this leads to considerable smoothing of $\Delta({\bm k}_F)$ \cite{Abrikosov1988metalsBOOK}. Indeed, in such materials $\Delta({\bm k}_F)$ is nearly constant along the Fermi surface, the strong possible anisotropy of the latter notwithstanding. In this case, commonly called ``$s-$wave", $\Delta$ is taken as a constant at the   Fermi surface:
\begin{equation}
       (\lambda^2)_{ik}^{-1}= \frac{8\pi  e^2 N(0)\langle   v_iv_k
\rangle}{ c^2}\,2\pi T\Delta^2  \sum_{\omega}  \frac{1
       }{\beta ^{3}}  \,.  \label{lambda-is}
\end{equation}
We obtain   in the Ginzburg-Landau domain:
\begin{equation}
  (\lambda^2)_{ik}^{-1}(T\to T_c)= 2(\lambda^2)_{ik}^{-1}(0)\,\,(1-T/T_c)\,,\qquad t=T/T_c  \,.
\label{lambda-isGL}
\end{equation}
At low temperatures:
\begin{eqnarray}
   \frac{\Delta (T)}{T_c} =\frac{\Delta (0)}{T_c} - \sqrt{\frac{2\pi T
  \Delta (0)}{T_c^2} } \,e^{- \Delta(0) /T}
 \approx   1.764  - 3.329
\sqrt{t}\, e^{-1.764/t}\,.
\label{D(T)}
\end{eqnarray}
The low-temperature behavior of the penetration depth is given by:
\begin{equation}
  (\lambda^2)_{ik}^{-1}= (\lambda^2)_{ik}^{-1}(0)\,
\left(1-2\sqrt{\frac{\pi\Delta(0)
       }{ T }}e^{- \Delta(0)/T} \right)  \,.
       \label{eq37}
\end{equation}

\subsubsection{2D $d-$wave: }

As an example of anisotropic $\Delta$, let us take a relatively simple but important case of a $d-$wave order parameter on the two-dimensional cylindrical Fermi surface: $\Omega = \Omega_0\cos 2\varphi$ where $\varphi$ is the properly chosen azimuth angle on the Fermi cylinder. With this choice, the gap nodes are at $\varphi=\pm \pi/4, \pm 3\pi/4$. The normalization Eq.(\ref{norm}) gives $\Omega_0= \sqrt{2}$, so that $\Delta_0 =\Psi\sqrt{2}\cos 2\varphi$. The order parameter at $T=0$ is now given by:
\begin{equation}
  \frac{\Delta_{max}(0)}{T_c} =   \frac{2\pi}{e^{ \gamma+0.5}} \approx
2.139\,.
\label{d-ratio}
\end{equation}
and at $T \ll T_c$:
\begin{equation}
   \frac{\Delta_{max}(T)}{T_c}  \approx  2.139  - 0.927\, t^3\,.
\label{****}
\end{equation}
After averaging over Fermi cylinder we obtain for the in-plane penetration depth:
\begin{eqnarray}
  \lambda ^{-2}(T) &=& \lambda ^{-2}(0)
\left(1-\sqrt{2}\,\frac{T}{\Delta_m}\right)\,,\label{lambda-d}\\
\lambda ^{-2}(0) &=&\frac{4\pi  e^2 N(0)v^2}{c^2 }\,   .
\label{lambda-d0}
\end{eqnarray}

\subsection{Eilenberger two-gap Scheme: The $\gamma$ - model}
\label{SEC:gamma-model}

 The full-blown microscopic approach based on the Eliashberg theory  is quite involved and not easy for the data analysis \cite{Golubov1997impurities,Brinkman2002mgB2Eliashberg,Dolgov2005MgB2,Nicol2005twoGaps}. Hence, the need for a relatively simple but justifiable, self-consistent and effective scheme experimentalists could employ. The weak-coupling model is such a scheme. Over the years, the weak-coupling  theory had proven to describe well multitude of superconducting phenomena. Similar  to the weak coupling is the ``renormalized BCS" model \cite{Nicol2005twoGaps} that incorporates the Eliashberg corrections in the effective coupling constants.  We call our approach a ``weak-coupling two-band scheme" and refer the reader to original papers where it is clarified that the applicability of the model for the analysis of the superfluid density and specific heat data is broader than the traditional weak coupling \cite{Kogan2002anisotropy,Kogan2009gamma}.

The Eilenberger approach can be used to describe self-consistently two-gap situation, in which
\begin{equation}
\Delta ({\bf k})= \Delta_{1,2}\,,\quad {\bf k}\in   F_{1,2} \,,
 \label{e50}
\end{equation}
  where $F_1,F_2$ are two separate sheets of the Fermi surface. Denoting the densities of states on the two parts as $N_{1,2}$, we have for a quantity $X$ constant at each Fermi sheet:
 \begin{equation}
\langle X \rangle = (X_1 N_1+X_2 N_2)/N(0) = n_1X_1+n_2X_2\,,
\label{norm2}
\end{equation}
where $n_{1,2}= N_{1,2}/N(0)$; clearly, $n_1+n_2=1$.

The  self-consistency equation (\ref{eil4}) now takes the form:
\begin{eqnarray}
\Delta_\nu= 2\pi T\sum_{ \mu=1,2} n_\mu \lambda_{\nu\mu}
f_\mu
=  \sum_{\mu} n_\mu \lambda_{\nu\mu}
\Delta_\mu \sum_{\omega }^{\omega_D}\frac{2\pi T}{\beta_\mu} ,
   \label{self-cons1}
\end{eqnarray}
where $\nu=1,2$ is the band index and $\lambda_{\nu\mu} = N(0)V(\nu,\mu)$ are dimensionless effective interaction constants. Note that the notation commonly used   in literature for $\lambda^{(lit)}_{\nu\mu}$ differs from ours: $\lambda^{(lit)}_{\nu\mu}=n_\mu \lambda_{\nu\mu}$.

Turning to the evaluation of $\Delta_\nu(T)$, we note that the sum over $\omega$ in Eq.\,(\ref{self-cons1}) is logarithmically divergent.  To deal with this difficulty, we employ  Eilenberger's idea of replacing $\hbar\omega_D$ with the measurable $T_c$.  Introducing dimensionless quantities
\begin{eqnarray}
\delta_\nu=  \frac{\Delta_\nu}{2\pi T} = \frac{\Delta_\nu}{ T_c}\, \frac{1}{2\pi t}\,,
   \label{deltas}
\end{eqnarray}
with $t=T/T_c$, we obtain:
\begin{eqnarray}
 \delta_\nu= \sum_{ \mu=1,2} n_\mu \lambda_{\nu\mu} \delta_\mu \left(  \frac{1}{{\tilde\lambda}}+\ln\frac{T_c}{T}
-A_\mu\right)\,, \nonumber\\ A_\mu =  \sum_{n=0}^{\infty}\left(\frac{1}{ n+1/2  }- \frac{1} {\sqrt{\delta_\mu^2+(n+1/2)^2}} \right)\,.
   \label{.pdf}
\end{eqnarray}
 where $\tilde\lambda$ is defined as:
\begin{eqnarray}
1.76\,T_c=  2\hbar\omega_D \exp (-1/{\tilde\lambda})\,,
   \label{Tc}
\end{eqnarray}
or
\begin{eqnarray}
\frac{1}{{\tilde\lambda}}=  \ln\frac{T}{T_c}+\sum_{\omega }^{\omega_D}\frac{2\pi T}{\hbar\omega}\,.
   \label{Eil0}
\end{eqnarray}
In terms of $\lambda_{\nu\mu}$, $\tilde\lambda$ is expressed as:
\begin{equation}
\tilde\lambda=\frac{2n_1n_2( \lambda_{11}  \lambda_{22}-\lambda_{12}^2
)}{ n_1 \lambda_{11}   + n_2 \lambda_{22} - \sqrt{ (n_1 \lambda_{11} - n_2 \lambda_{22})^2 -4n_1n_2\lambda^2_{12}}}\,.
 \label{S1}
\end{equation}

\begin{figure}[h]
\centering
\includegraphics[width=9cm]{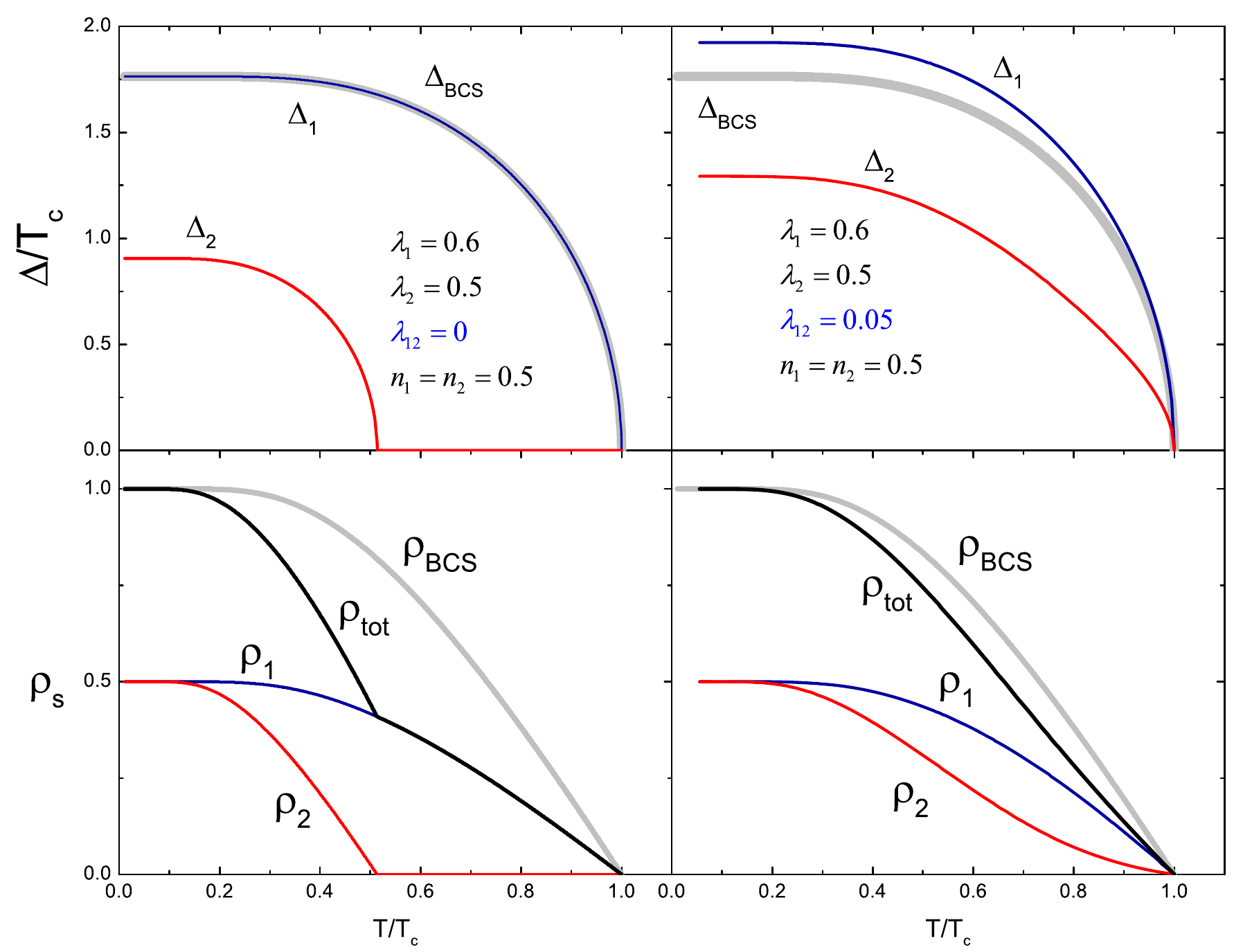}%
\caption{Top row: self-consistent gaps calculated for the indicated $\lambda_{\nu\mu}$ in a compensated metal with $\mu = 1$. First graph is for no inter-band pairing, $\lambda_{12}=0$. The second for $\lambda_{12}=0.05$. Bottom row: corresponding partial and total superfluid densities.}
\label{diffLambdas}
\end{figure}

For the given coupling constants $\lambda_{\nu\mu}$ and densities of states $n_\nu$, this system  can be solved numerically for $\delta_\nu$ and therefore provide the gaps $\Delta_\nu=2\pi T\delta_\nu(t)$. Example calculations are shown in Fig.~\ref{diffLambdas}. First graph in the top row is calculated assuming no interband coupling. Naturally, we obtain material with two different transition temperatures. The second graph shows the gaps with $\lambda_{12}=0.05$, which features a single $T_c$ and quite non-single-BCS-gap temperature dependence of the smaller gap. The single BCS gap is shown for comparison.

\paragraph{Superfluid density: }

Having formulated the way to evaluate $\Delta(T)$, we turn to the London  penetration depth  given for general anisotropies of the Fermi surface and of $\Delta$ by Eq.~(\ref{lambda-tensor}) \cite{Kogan2002anisotropy}. We consider here only the case of currents in the $ab$ plain of uniaxial or cubic materials with two separate Fermi surface sheets, for which the superfluid density,
$\rho=\lambda_{ab}^2(0)/\lambda_{ab}^2(T)$, is:

\begin{eqnarray}
\rho  &=& \gamma {\rho _1} + \left( {1 - \gamma } \right){\rho_2}\,,
\nonumber\\
\rho _\nu &=&  \delta_\nu^2\sum_{n=0}^\infty
  \left[\delta_\nu^2+(n+1/2)^2\right]^{-3/2}\,,
\nonumber\\
\gamma  &=& \frac{{{n_1}v_{1}^2}}{{{n_1}v_{1}^2 +
{n_2}v_{2}^2}}\,.
\label{rhogamma}
\end{eqnarray}

\noindent where $v_{\nu}^2$ are the averages of the in-plane Fermi velocities over the corresponding band.

With the discovery of two-gap superconductivity in a number of materials, including
MgB$_2$ \cite{Bouquet2001CpMgB2,Fletcher2005MgB2}, Nb$_2$Se \cite{Fletcher2007NbSe2}, V$_3$Si \cite{Nefyodov2005V3Si}, Lu$_2$Fe$_3$Si$_5$ \cite{Gordon2008Lu2Fe3Si5} and ZrB$_{12}$ \cite{Gasparov2006ZrB12} one of the most popular approaches to analyze the experimental results has been the so called ``$\alpha$-model" \cite{Bouquet2001CpMgB2}. Developed originally to renormalize a single weak-coupling BCS gap to account for strong - coupling corrections \cite{Padamsee1973StrongCoupling}, it was used to introduce two gaps, $\Delta_{1,2}$, each having a BCS temperature dependence, but different amplitudes \cite{Bouquet2001CpMgB2}. This allowed for a simple way to fit the data on the specific heat \cite{Bouquet2001CpMgB2} and the superfluid density, $\rho=x\rho_1+(1-x)\rho_2$ \cite{Fletcher2005MgB2,Prozorov2006SUSTreview}. Here, $\rho_{1,2}$ are evaluated with $\Delta_{1,2}=(\alpha_{1,2}/1.76)\Delta_{BCS}(T)$ and $x$ takes into account the relative band contributions. The fitting is usually quite good (thanks to a smooth and relatively ``featureless'' $\rho(T)$ and the parameters $\alpha$ where found to be one larger and one smaller than unity (unless they are both equal to one in the single-gap limit). Although the $\alpha$-model had played an important and timely role in providing convincing evidence for the  two-gap superconductivity, it is \emph{intrinsically inconsistent} as applied to the full temperature range. The problem is that one cannot assume \textit{a priory} temperature dependencies for the gaps in the presence of however weak interband coupling (required to have single $T_c$). In unlikely situation of zero interband coupling, two gaps would have single-gap BCS-like $T-$ dependencies, but will have two different transition temperatures. The formal similarity in terms of additive partial superfluid densities prompted to name our scheme the ``$\gamma$-model". We note, however, that these models are quite different: our $\gamma$ that determines partial contributions from each band is not just a partial density of states $n_1$ of the $\alpha$-model, instead it involves the band's Fermi velocities. The gaps, $\Delta_{1,2}(T)$, are calculated self-consistently during the fitting procedure.

\begin{figure}[tbh]
\centering
\includegraphics[width=9cm]{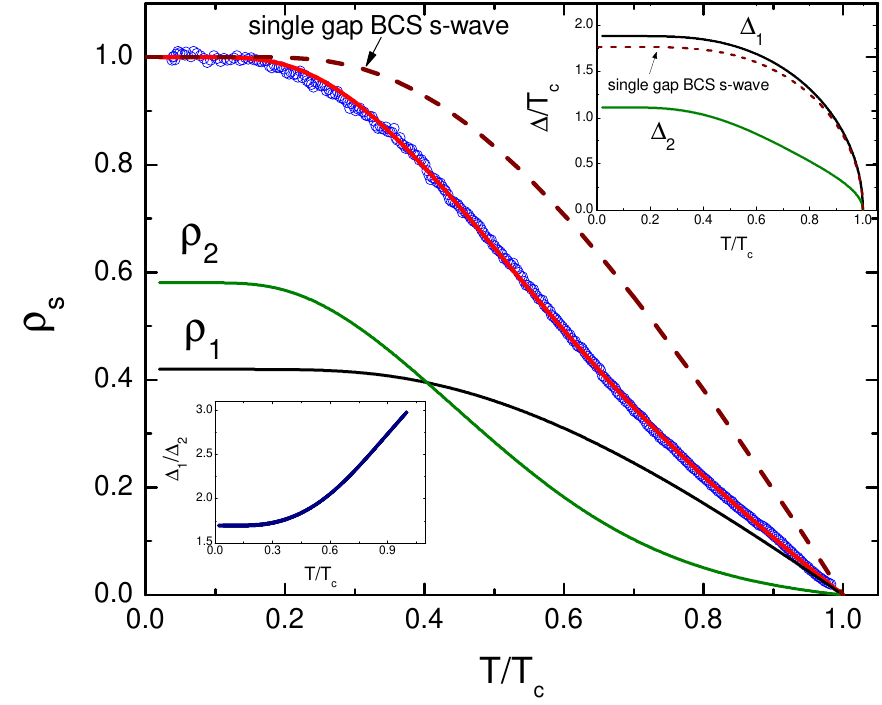}%
\caption{\label{fig4}Symbols: superfluid density, $\rho_s(T)$ calculated with $\lambda(0)=200$ nm. Solid lines represent the fit to a two-gap $\gamma-$model, $\rho_s = \gamma \rho_1 + (1-\gamma) \rho_2$. Dashed line is a single-gap BCS solution. Upper inset: superconducting gaps, $\Delta_1(T)$ and $\Delta_2(T)$ calculated self-consistently during the fitting. Lower inset: $\Delta_1/\Delta_2$ as a function of temperature.}
\label{LiFeAs}
\end{figure}

The $\gamma$-models can be simplified for a compensated metal, such as clean stoichiometric superconductor LiFeAs \cite{Kim2011LiFeAsLambda}. To reduce the number of fitting parameters, yet capturing compensated multiband structure, we consider a simplest model of two cylindrical bands with the mass ratio, $\mu = m_1/m_2$, whence the partial density of states of the first band, $n_1=\mu/(1+\mu)$. The total superfluid density is $\rho_s = \gamma \rho_1 + (1-\gamma) \rho_2$ with  $\gamma=1/(1+\mu)$. We also use the Debye temperature of 240 K \cite{Wei2010CpMultigap} to calculate $T_c$, which allows fixing one of the in-band pairing potentials, $\lambda_{11}$. This leaves three free fit parameters: the second in-band potential, $\lambda_{22}$, inter-band coupling, $\lambda_{12}$, and the mass ratio, $\mu$. Indeed, we found that $\rho_s(T)$ can be well described in the entire temperature range by this clean-limit weak-coupling BCS model \cite{Kim2011LiFeAsLambda}. Figure \ref{LiFeAs} shows the fit of the experimental superfluid density to the $\gamma-$model. The inserts show temperature - dependent superconducting gaps obtained as a solution of the self-consistency equation, Eq.~(\ref{self-cons1}) and the lower inset show the gap ration as a function of temperature. Evidently, the smaller gap does not exhibit a BCS temperature dependence emphasizing the failure of the commonly used $\alpha-$model.

Lastly we note that in order to have two distinct gaps as observed in many experiments \cite{Johnston2010review} one has to have significant in-band coupling constants, $\lambda_{11}$ and $\lambda_{11}$. To support this conclusion we calculate the gap ratio $\gamma_{\Delta}=\Delta_1(0)/\Delta_2(0)$ for fixed $\lambda_{12}$ and varying $\lambda_{11}$ and $\lambda_{22}$. Red lines in each graph show $\gamma_{\Delta}=2$. For all cases, one needs significant $\lambda_{12}$ and varying $\lambda_{11}$ to reach this gap ratio. Therefore, the original simplified $s_{\pm}$ model with two identical Fermi surfaces and only interband coupling, $\lambda_{12} \neq 0$ and $\lambda_{11}=\lambda_{22}=0$ does not describe the experimentally found two distinct gaps in Fe-based superconductors \cite{Mazin2008spm}.

\begin{figure}[h]
\centering
\includegraphics[width=9cm]{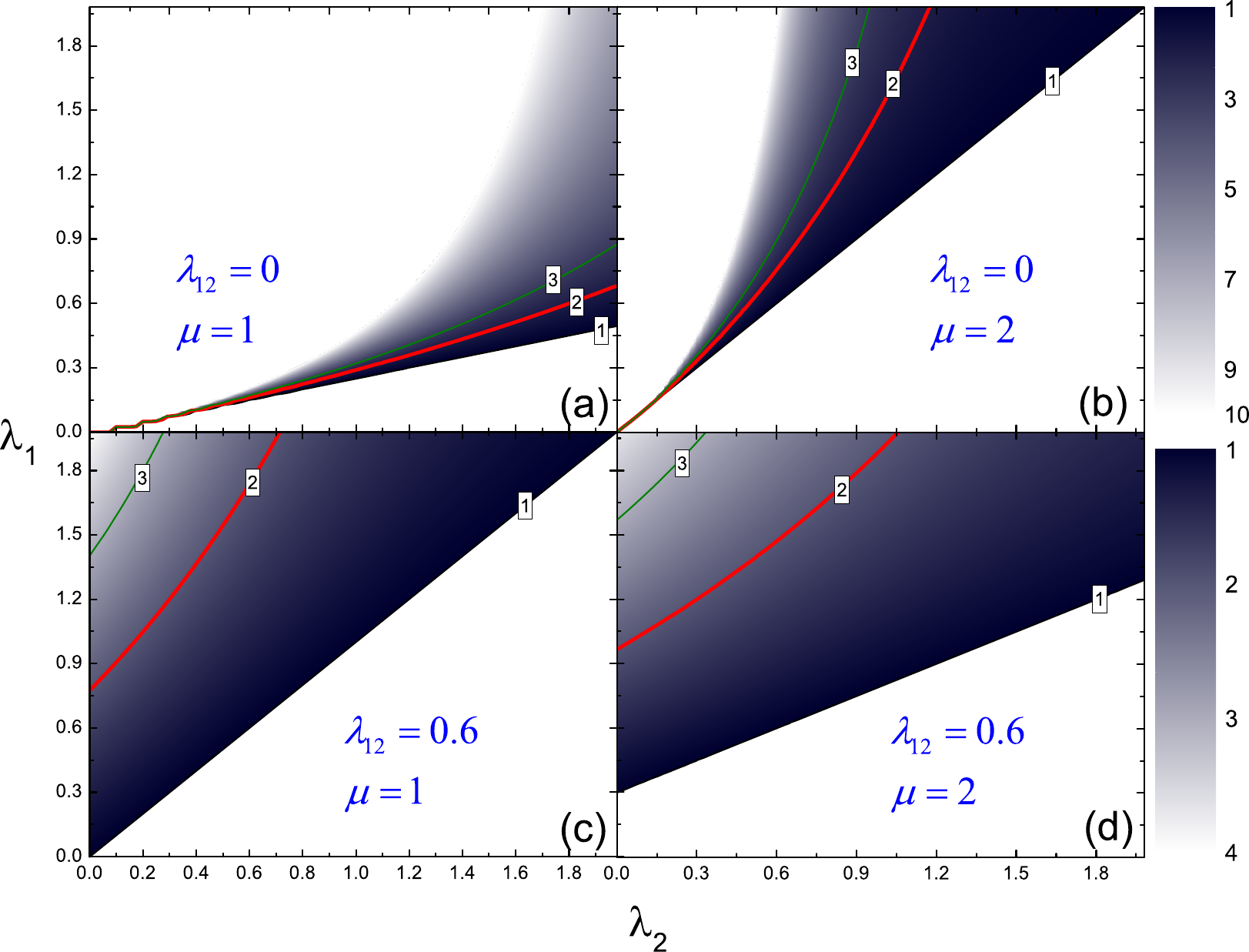}%
\caption{Gap ratio, $\gamma_{\Delta}=\Delta_1(0)/\Delta_2(0)$ for indicated $\mu$ (to check the effect of different partial densities of states) and $\lambda_{12}$. The magnitude of $\gamma_{\Delta}$ is given by the intensity shown on the right. Red lines in each graph show $\gamma_{\Delta}=2$. }
\label{GapRatio}
\end{figure}

\section{Effects of Scattering}\label{SEC:pairbreaking}

\paragraph{}The scattering by impurities strongly affects the London penetration depth even in the simplest case of non-magnetic impurities in materials with isotropic gap parameter and for scattering processes which can be characterized by the scalar (isotropic) scattering rate. The $\Delta$ anisotropy and magnetic impurities complicate the analysis of
$\lambda(T)$, among other reasons due to the $T_c$ suppression in these
cases. For extended treatment the reader is referred to \cite{Kogan2009pairBreaking,Gordon2010Pairbreaking} and here we only summarize properties related to the discussion of our results. Introducing the non-magnetic scattering rate, $1/\tau$, and magnetic scattering rate (spin - flip), $1/\tau_m$, London penetration depth is expressed as:

\begin{equation}
(\lambda^2)_{ik}^{-1}= \frac{16\pi^2 e^2TN(0)
\langle v_iv_k\rangle}{
c^2}\,  \sum_{\omega}  \frac{f_0^2
}{\Delta/f_0+\hbar/2\tau^-}  \,.  \label{lambda-m}
\end{equation}

\noindent here $1/\tau^{\pm} = 1/\tau \pm 1/\tau_m$. Clearly, the evaluation of $\lambda(T)$ in the presence of magnetic impurities is quite involved. There is one limit, however, for which we have a simple analytic answer:

\paragraph{Gapless limit:}

This is the case when $\tau_m$ is close to $ 2\hbar/\Delta_0(0)$, the critical value for which $T_c=0$, i.e. $\tau_m\Delta \ll 1$.
The resulting expression for the order parameter is remarkably simple \cite{AbrikosovGorkov1960}:

\begin{equation}
\Delta^2 = 2\pi^2(T_c^2-T^2)
\label{AG_Delta}
\end{equation}

The result for superfluid density is valid in the entire temperature domain, $T<T_c$.

\begin{equation}
(\lambda^2)_{ik}^{-1}= \frac{8\pi  e^2N(0)
\langle v_iv_k\rangle}{
c^2 (\rho^-)^2}\left(    \ln\frac{2\rho_m}{\rho^+}+\frac{\rho^-}{2\rho_m
}\right)(1-t^2).  \label{lambda-gapless}
\end{equation}

\noindent where $\rho=\hbar/(2\pi T_c \tau)$, $\rho_m=\hbar/(2\pi T_c \tau_m)$ and $\rho^{\pm}=\rho \pm \rho_m$. For a short transport mean-free path $\rho\gg\rho_m$ we have
Abrikosov-Gor'kov's result:

\begin{equation}
(\lambda^2)_{ik}^{-1}= \frac{8\pi^3 e^2N(0)
\langle v_iv_k\rangle}{
c^2  \rho\rho_m}\,(1-t^2).  \label{lambda-gaplessAG}
\end{equation}

The idea of strong pair-breaking and, perhaps, gapless superconductivity finds experimental evidence in Fe-based superconductors in form of scaling relations for the specific heat jump and a pre-factor of the quadratic temperature variation of $\lambda(T)$ \cite{Kogan2009pairBreaking,Gordon2010Pairbreaking}.

To summarize, in the case of superconductor with line nodes, impurity scattering will change linear temperature dependence of the penetration depth at $T \ll T_c$ to become quadratic \cite{Hirschfeld1993dwave}. Disorder will also lift the c-axis line nodes in case of extended $s-$wave and induce a change from effective $T^2$ to exponentially activated behavior for the in-plane penetration depth \cite{Mishra2009nodesLifting}. However, if we start in the clean limit of a fully - gapped superconductor and introduce pair-breaking scattering (either due to magnetic impurities or due to unconventional gap structure, such as $s_{\pm}$) \cite{Mazin2008spm}, penetration depth will also exhibit a power-law behavior approaching $T^2$ variation in the gapless limit. Therefore, with increasing impurity scattering, if $\lambda(T) \sim T^n$, we expect the exponent $n$ to change from 1 to 2 in case of nodal superconductor (or even from 1 to $\sim\exp$ in case of non symmetry - imposed nodes) and from $\sim\exp$ to 2 in case of fully-gapped material with pair-breaking. This is schematically illustrated by shaded areas in Fig.~\ref{n_vs_Tc} (where we used large values of the exponent $n$ to designate $\sim\exp$ behavior).

\section{Experimental results}\label{SEC:DATA}

\begin{figure}[h]
\centering
\includegraphics[width=9cm]{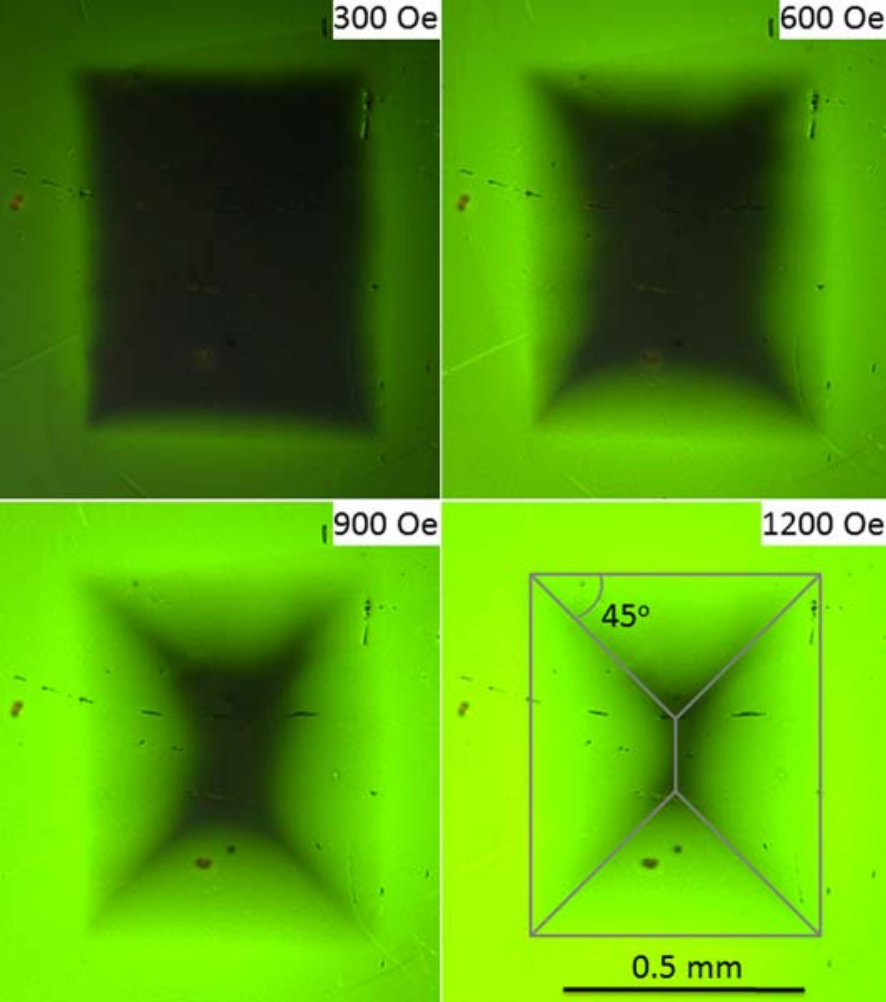}%
\caption{Magnetic flux penetration at 20 K into a crystal with x=0.07. The last frame shows a schematic overlay of the expected "Bean oblique wedge" shape with isotropic in-plane current density.}
\label{FeCo122MO1}
\end{figure}

\begin{figure}[h]
\centering
\includegraphics[width=9cm]{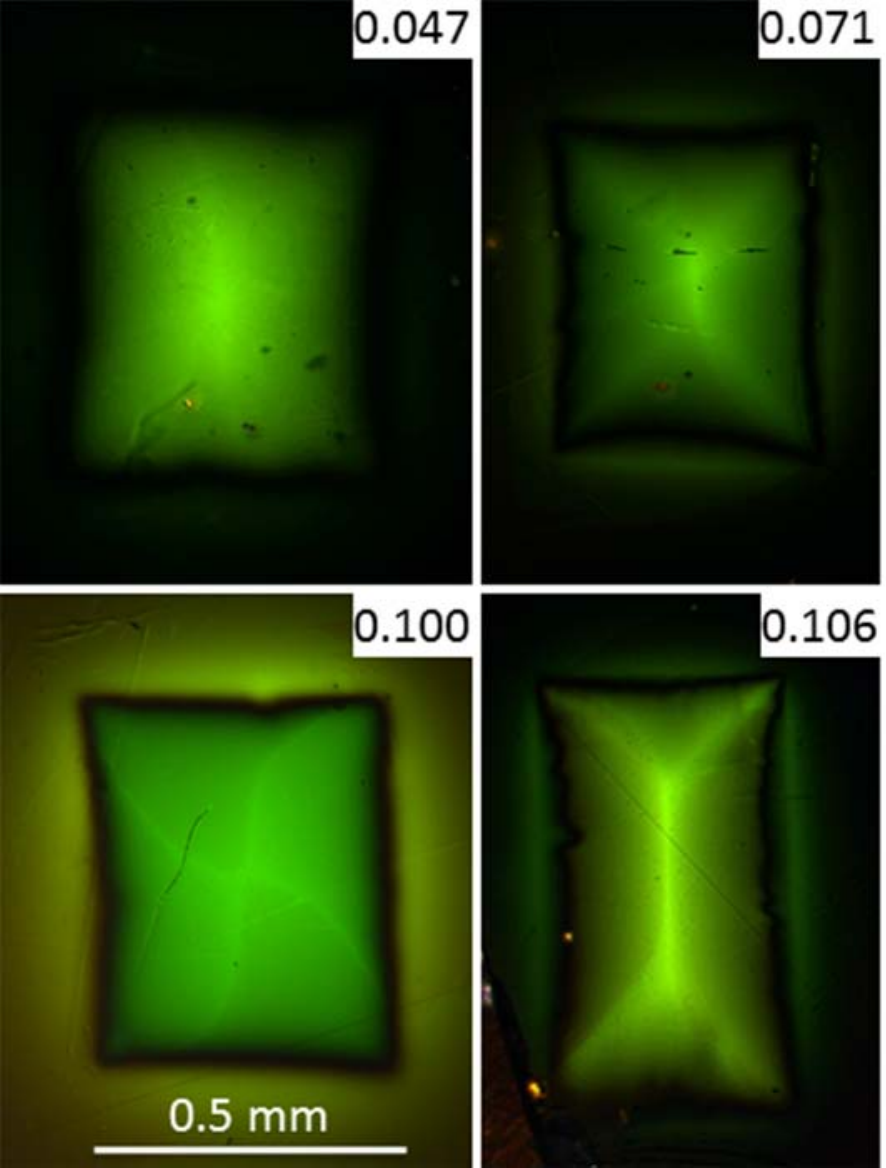}%
\caption{Magnetic flux trapped in samples after cooling in a 1.5 kOe magnetic field to 5 K and turning field off. Doping levels are indicated in right top corners.}
\label{FeCo122MO2}
\end{figure}

\paragraph{}
Ba(Fe$_{1-x}$T$_x$)$_2$As$_2$ is one of the most studied systems among all Fe-based superconductors and we have collected extensive data that illustrate general features often common to many other members of the diverse pnictide family. Here we focus on the electron doped Ba(Fe$_{1-x}T_x$)$_2$As$_2$ with $T=$ Co and Ni.  One reason why these series were chosen is because large, high quality single crystals are available \cite{Canfield2010review122}.
All samples were grown from the self-flux and were extensively characterized by transport, structural, thermal and magneto-optical analysis. They all exhibited uniform superconductivity at least at the 1 $\mu m$ scale and dozens of samples were screened before entering into the resulting discussion \cite{Ni2008FeCoHc2,Canfield2010review122}. To demonstrate sample quality we show magneto-optical images in Fig.~\ref{FeCo122MO1} and Fig.~\ref{FeCo122MO2}. Details of this visualization technique are described elsewhere \cite{Prozorov2009vortices}. In the images intensity is proportional to the local magnetic induction. All samples show excellent Meissner screening \cite{Prozorov2009vortices}. Fig.~\ref{FeCo122MO1} shows penetration of the magnetic field into the optimally doped sample at 20 K. A distinct "Bean oblique wedge" shape \cite{Bean1964} of the penetrating flux with the current turn angle of 45$^o$ (implying isotropic in-plane current density) is observed.

Furthermore, to look for possible mesoscopic faults and inhomogeneities, we show the trapped magnetic flux obtained after cooling in a 1.5 kOe magnetic field and turning field off. The vortex distribution is quite homogeneous indicating robust uniform superconductivity for various doping levels. This is shown in Fig.~\ref{FeCo122MO1} for four different doping levels.

The parent compound, BaFe$_2$As$_2$ is a poor metal \cite{Tanatar2009parent122} having a high temperature tetragonal phase with no long range magnetic order and undergoes structural and magnetic transitions around 140 K into a low temperature orthorhombic phase with long range antiferromagnetic spin density wave order \cite{Canfield2010review122}. Transition metal doping onto the iron site serves to suppress the structural and magnetic transition temperatures and superconductivity emerges after magnetism has sufficiently been weakened. Doping into barium site with potassium results in hole doped superconductivity. Properties of this hole - doped system, at least as far as penetration depth is concerned, are quite similar to the electron - doped FeT122 \cite{Hashimoto2009BaK122,Martin2009BaK}. On the other hand, properties materials obtained by isovalent doping of phosphorus into the arsenic site seems to induce superconductivity without introducing significant scattering and seems to result in a superconducting gap with line nodes \cite{Hashimoto2010AsPnodesPD}.

In the charge - doped systems, angle-resolved spectroscopy (ARPES) \cite{Ding2008ARPESBaK122,Evtushinsky2009ARPESBaK,Liu2009ARPES,Xu2011BaK3DgapARPES}
 and thermal conductivity \cite{Reid2010FeCo122TC,Tanatar2010FeCoTCvsX} consistently show fully gapped Fermi surfaces (however with gap anisotropy increasing upon departure from optimal doping in case of thermal conductivity \cite{Tanatar2010FeCoTCvsX}.) Furthermore, the $c-$axis behavior is quite different and it is possible that a nodal state develops upon doping beyond optimal level \cite{Martin2010FeNinodes,Reid2010FeCo122TC}. The in-plane penetration depth consistently shows non-exponential power-law behavior \cite{Bobowski2010MW,Gofryk2010Cp,Gordon2010Pairbreaking,Gordon2009FeCo122vsX,Gordon2009FeCo122Opt,Hashimoto2009BaK122,Kim2010irr,Luan2010FeCo122Opt,Luan2011FeCoL0,Martin2009BaK,Martin2010SUST,Prozorov2010SUST,Prozorov2009physC,Williams2010FeCouSR,Williams2009uSRL0}, which will be discussed in detail below. It seems that such behavior can be explained by the pair-breaking scattering \cite{Bang2009,Dolgov2009,Senga2008,Vorontsov2009b,Mishra2011}. This is supported experimentally by the observed variation of the low-temperature $\lambda(T)$ within nominally the same system (and even in pieces of the same sample) \cite{Hashimoto2009BaK122} as well as deliberately introduced defects \cite{Kim2010irr}. In this review we also discuss the case of a substantial variation of $\lambda(T)$ between various samples, probably due to edge effect.

In the following analysis we use two ways to represent the power law behavior: $\lambda(T) = \lambda(0) + A(T/T_{c})^2$ at low temperatures (below 0.3 $T_c$) with $A$ being the only free parameter, because at a gross level, all samples follow the $\lambda(T) \sim T^{2}$ behavior rather well and $\lambda(T) = \lambda(0) + CT^{n}$, leaving the exponent $n$ as free parameter to analyze its evolution with doping or artificially introduced defects. In the case of vertical line nodes, we expect a variation from $n=1$ to $n=2$ upon increase of pair-breaking scattering \cite{Hirschfeld1993dwave}, but in the case of a fully gapped $s_{\pm}$ state we expect an opposite trend to approach $n=2$ in the dirty limit from clean - limit exponential behavior \cite{Bang2009,Dolgov2009,Senga2008,Vorontsov2009b}. If, however, nodes are formed predominantly along the c-axis in the extended $s-$wave scenario, the effect of scattering on the in-plane penetration depth would be opposite - starting from roughly $n=2$ in the clean limit and approaching exponential in the dirty limit \cite{Mishra2011}.

\subsection{In-plane London penetration depth}

\paragraph{}

\begin{figure}[h]
\centering
\includegraphics[width=9cm]{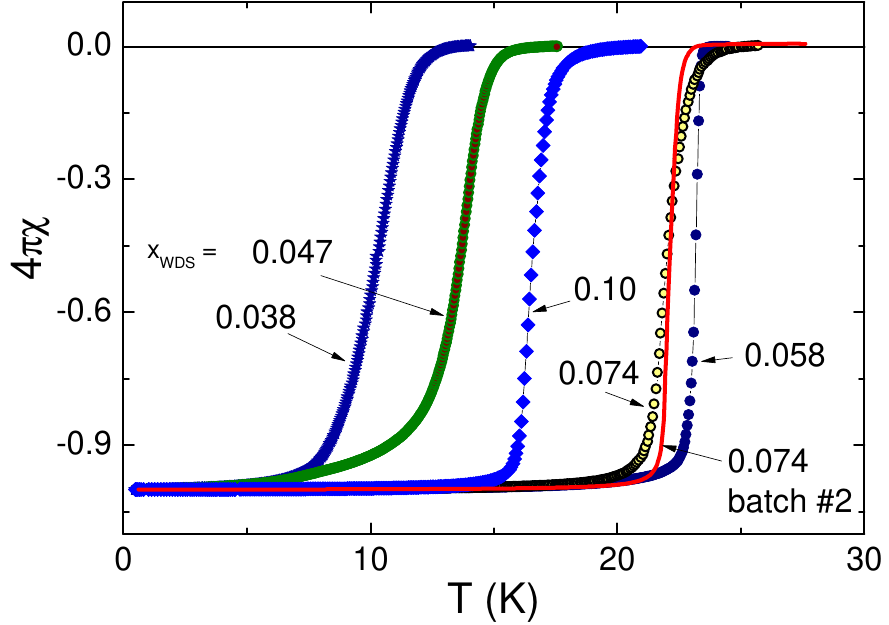}%
\caption{$4\pi\chi\left( T\right)$ in single crystals of Ba(Fe$_{1-x}$Co$_{x}$)$_{2}$As$_{2}$ for different $x$. Sample from batch \#1 is shown by a solid line.}
\label{FeCoFig1}
\end{figure}

Figure \ref{FeCoFig1} shows normalized differential TDR magnetic susceptibility of several single crystals of Ba(Fe$_{1-x}$Co$_x$)$_2$As$_2$ across the superconducting ``dome''. All but one samples were grown at the same conditions and with similar starting chemicals. All, but one had thicknesses in the range of 100 - 400 nm. One of the samples ($x=0.074$, denoted batch \#2) was cleaved for the irradiation experiments and had thickness of 20 nm. We use it for comparison with the ``thick'' batch \#1 and, also, to study the effects of deliberately induced defects. It turns out that the edges of the thicker samples are not quite smooth and, when imaged in SEM, look like a used book (see Fig.~\ref{FeCoFig5}(a)). Since calibration of the TDR technique relies on the volume penetrated by the magnetic field, the thinner samples should be closer to the idealization of the sample geometry (top and bottom surfaces are always very flat and mirror-like), thus producing a more reliable calibration. On the other hand, this would only lead to a change of the amplitude (due to geometric mis-calibration) of the penetration depth variation (i.e., pre-factor $A$) and would not change its functional temperature dependence (i.e., the exponent $n$). Thinner samples, on the other hand, have better chance to be more chemically uniform, thus have reduced scattering. We observed these effects comparing samples from different batches.

\begin{figure}[h]
\centering
\includegraphics[width=9cm]{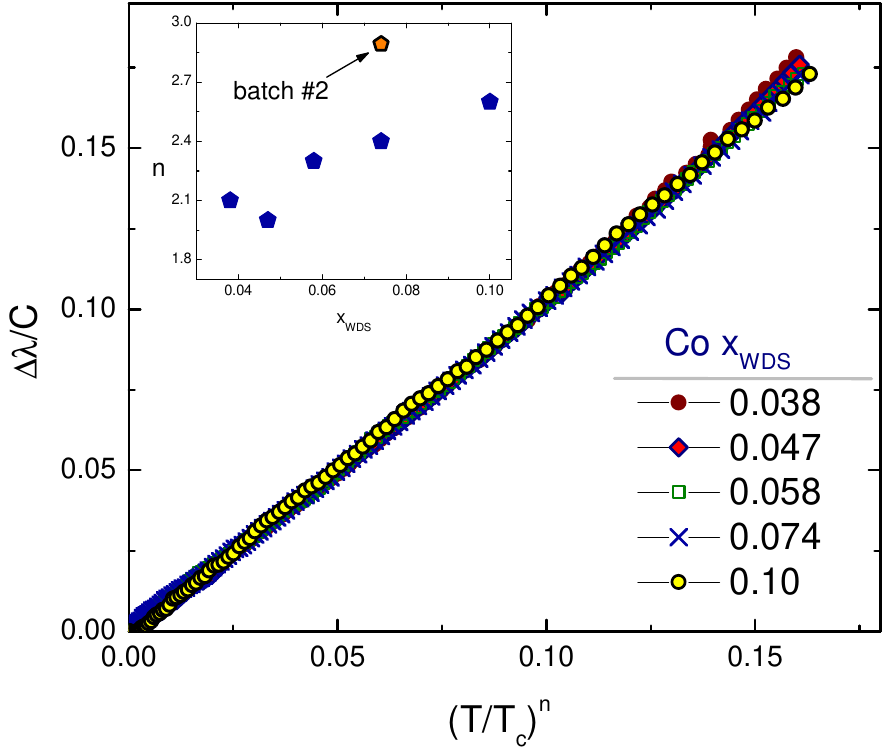}%
\caption{Low-temperature behavior of $\Delta \lambda(T)$ vs. $(T/T_c)^n$ for all
studied concentrations. Inset shows exponent $n$ as function of concentration. Sample from batch \#2 is shown for comparison.}
\label{FeCoFig2}
\end{figure}

Low-temperature variation of London penetration depth is presented in Fig.~\ref{FeCoFig2} as a function of $(T/T_c)^n$ obtained by fitting the data to $\Delta \lambda(T) = \lambda(0) + C(T/T_c)^n$. Each curve reveals a robust power law behavior with the exponent $n$ shown in the inset in Fig.~\ref{FeCoFig2}. The fitted exponent $n$ varies from $n = 2 \pm 0.1$ for underdoped samples to $n=2.5\pm0.1$ for the overdoped samples within the batch \#1 and reaches $n=2.83$ in batch \#2. If the superconducting density itself follows a power law with a given $n$, then $C \sim f_s(c/\omega_{p}) S$, where $f_{s}$ is the superconducting fraction at zero temperature, $c$ is the speed of light and $S$ is defined by the fraction of the Fermi surface that is gapless (which may reflect a multigap character of the superconductivity, possible nodal structure, unitary impurity scattering strength, etc) and $\omega_{p}$ is the plasma frequency.

\begin{figure}[h]
\centering
\includegraphics[width=9cm]{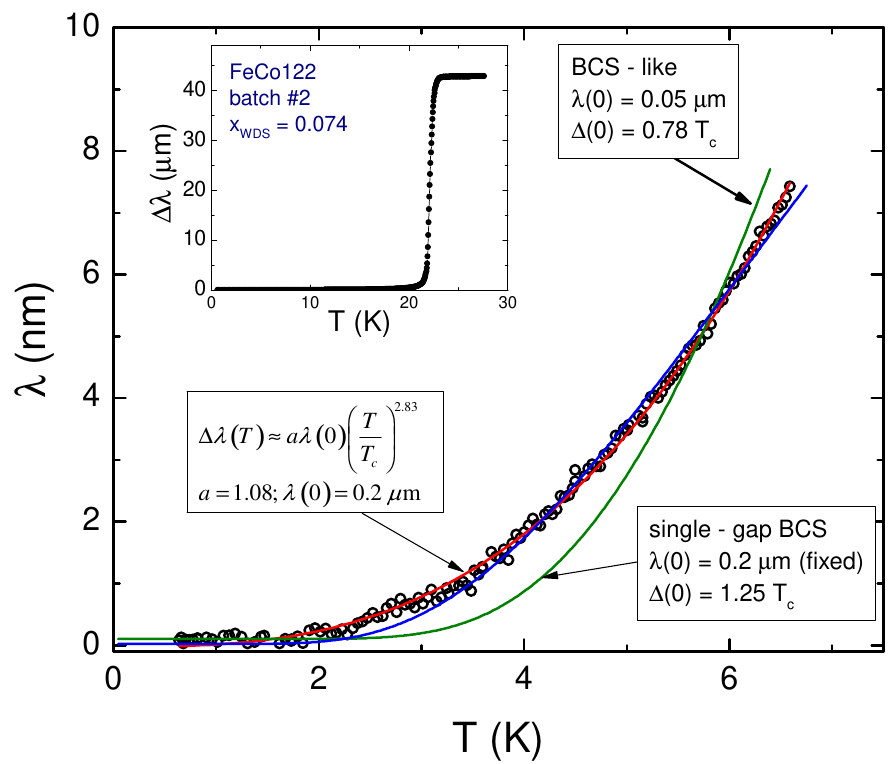}%
\caption{Low-temperature behavior of $\Delta \lambda(T)$ for the sample batch \#2. Solid lines are the fits to three functions described in the text. Insert shows full temperature variation indicating very sharp superconducting transition.}
\label{FeCoFig3}
\end{figure}

Clearly the sample of batch \#2 shows behavior much closer to exponential compared to batch \#1. As discussed above, this could be due to the variation of scattering between the batches. We analyze low-temperature $\lambda(T)$ in Fig.~\ref{FeCoFig3}. We attempted to fit the data with three functions: the power-law with free pre-factor $C$ and exponent $n$, the standard single-gap BCS behavior, Eq.~(\ref{eq37}), with a fixed value of $\lambda(0) = 200$ nm and $\Delta(0)$ as a free parameter and to a BCS - like function where both $\lambda(0)$ and $\Delta(0)$ were free parameters. The resulting values are shown in Fig.~\ref{FeCoFig3}.
The power-law fit yields quite high exponent $n \approx 2.83$ and the best fit quality. The BCS - like fit yields a reasonable fit quality, but produces impossible values of both $\lambda(0) \approx 48$ nm and $\Delta(0)\approx 0.78T_c$ (the latter cannot be less than a weak coupling BCS value of 1.76. Finally, the fixed $\lambda(0)$ BCS fit does not agree with the data and also produces unreasonable $\Delta(0)\approx 1.25T_c$. One strong conclusion follows from this exercise - we are dealing with a multi-gap superconductor.

\begin{figure}[h]
\centering
\includegraphics[width=9cm]{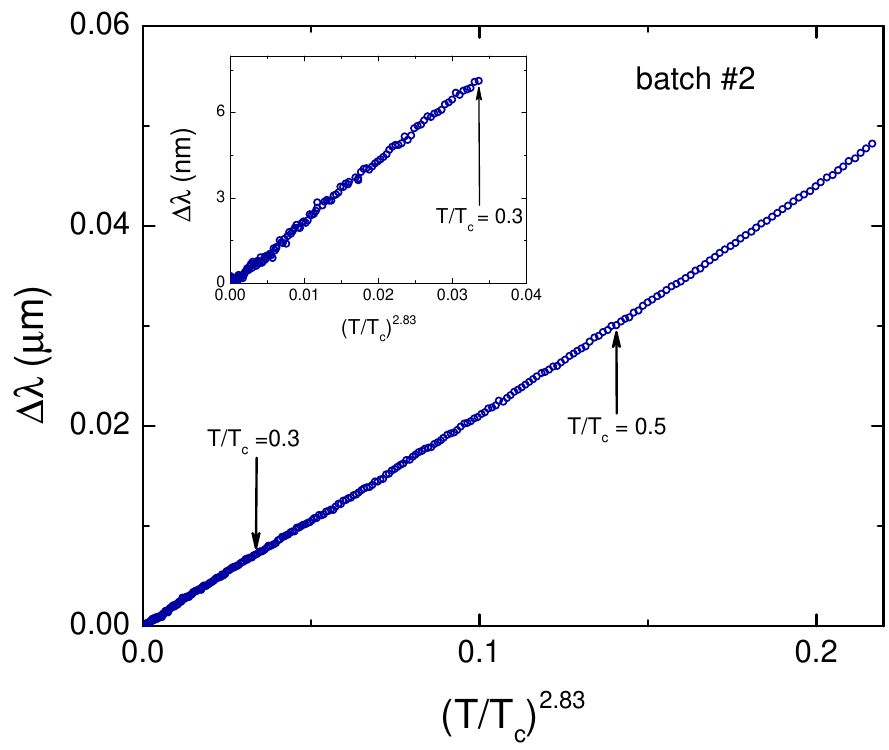}%
\caption{Low-temperature behavior of $\Delta \lambda(T)$ vs. $(T/T_c)^{2.83}$ for the sample from batch \#2. Marks show actual reduced temperature. Insets show the behavior below commonly accepted ``low-temperature limit'' of $\approx T_c/3$.}
\label{FeCoFig4}
\end{figure}

In order to understand the validity of the empirical power-law behavior, Fig.~\ref{FeCoFig4} shows low-temperature behavior of $\Delta \lambda(T)$ vs. $(T/T_c)^{2.83}$ for the sample from batch \#2. Arrows show actual reduced temperature. Inset zooms at below of the commonly accepted ``low-temperature limit'' of $\approx T_c/3$. Clearly, power-law behavior is robust and persists down to the lowest temperature of the experiment of $T \approx 0.02 T_c$.

\begin{figure}[h]
\centering
\includegraphics[width=12cm]{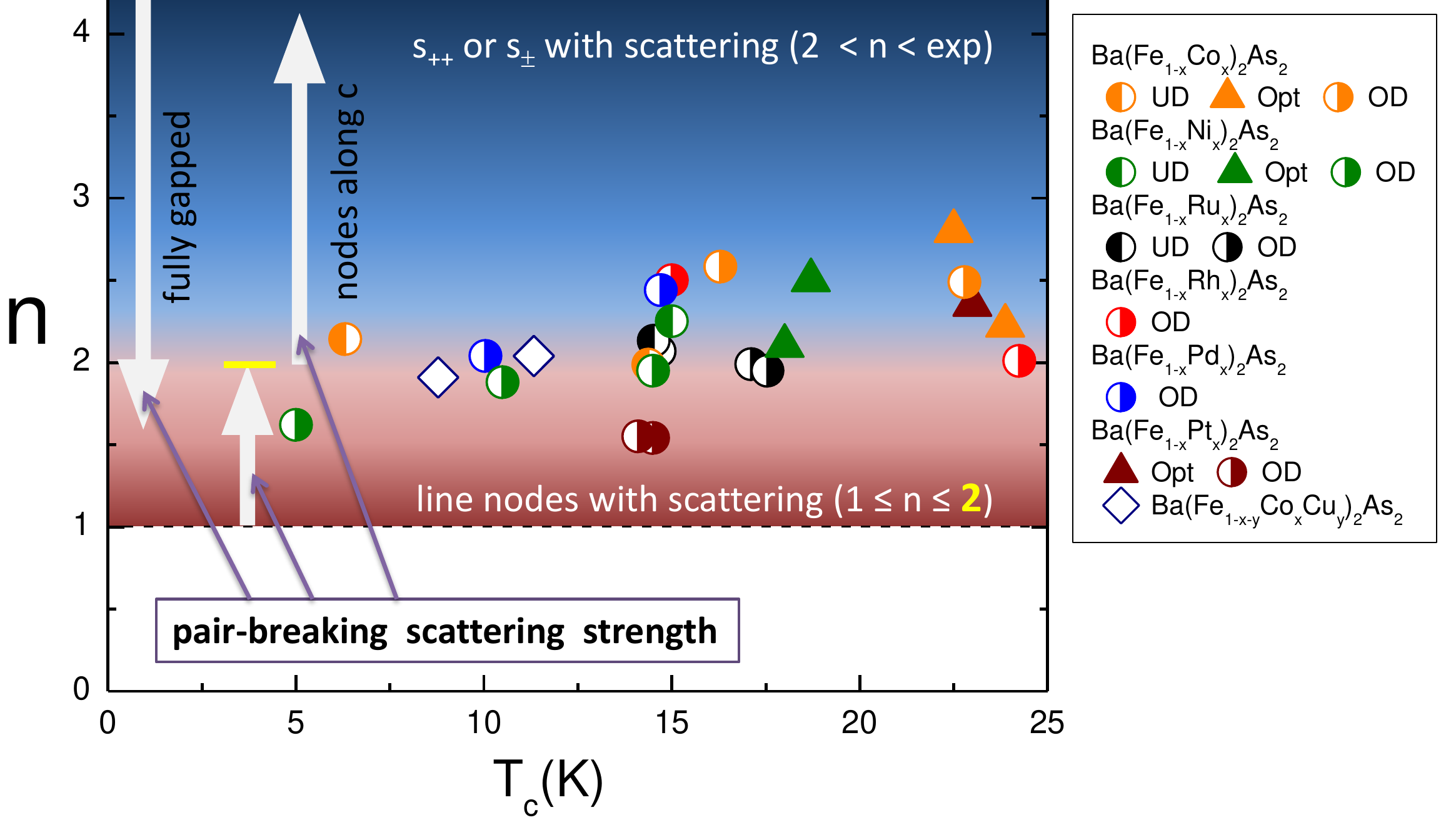}%
\caption{Power-law exponent of the low-temperature variation of in-plane $\lambda_{ab}(T)$ is several electron-doped Ba122 superconductors at various doping levels. Shaded areas show the influence of pair-breaking scattering with $n=2$ being the limiting value of $n$ approaching from either the $s-$wave side (nodeless exponential) or $d-$wave side (vertical line nodes). In case of extended $s-$wave with nodes predominantly along the c-axis, effective $n$ increasing from 2 in the clean limit towards the exponential behavior when the nodes are lifted by scattering.}
\label{n_vs_Tc}
\end{figure}

We now summarize the observed power-law behavior of the in-plane penetration depth for the electron - doped 122 family of superconductors. Figure \ref{n_vs_Tc} shows the experimental low-temperature limit power-law exponent for different dopants on the Fe site and at different doping regimes. The shaded areas show the expectations for the pair-breaking scattering effects in $s-$ and $d-$wave scenario. It seems that statistically $d-$wave pairing (more generally - vertical line nodes) cannot explain the in-plane variation of the penetration depth. However, if the nodes appear somewhere predominantly along the c-axis, they may induce an apparent power law behavior of $\lambda_{ab}(T)$ with the effective $n \approx 2$ in the clean limit and going towards exponential behavior with the increase of the scattering rate \cite{Mishra2011}.

\subsection{Absolute value of the penetration depth}

\paragraph{}

\begin{figure}[h]
\centering
\includegraphics[width=9cm]{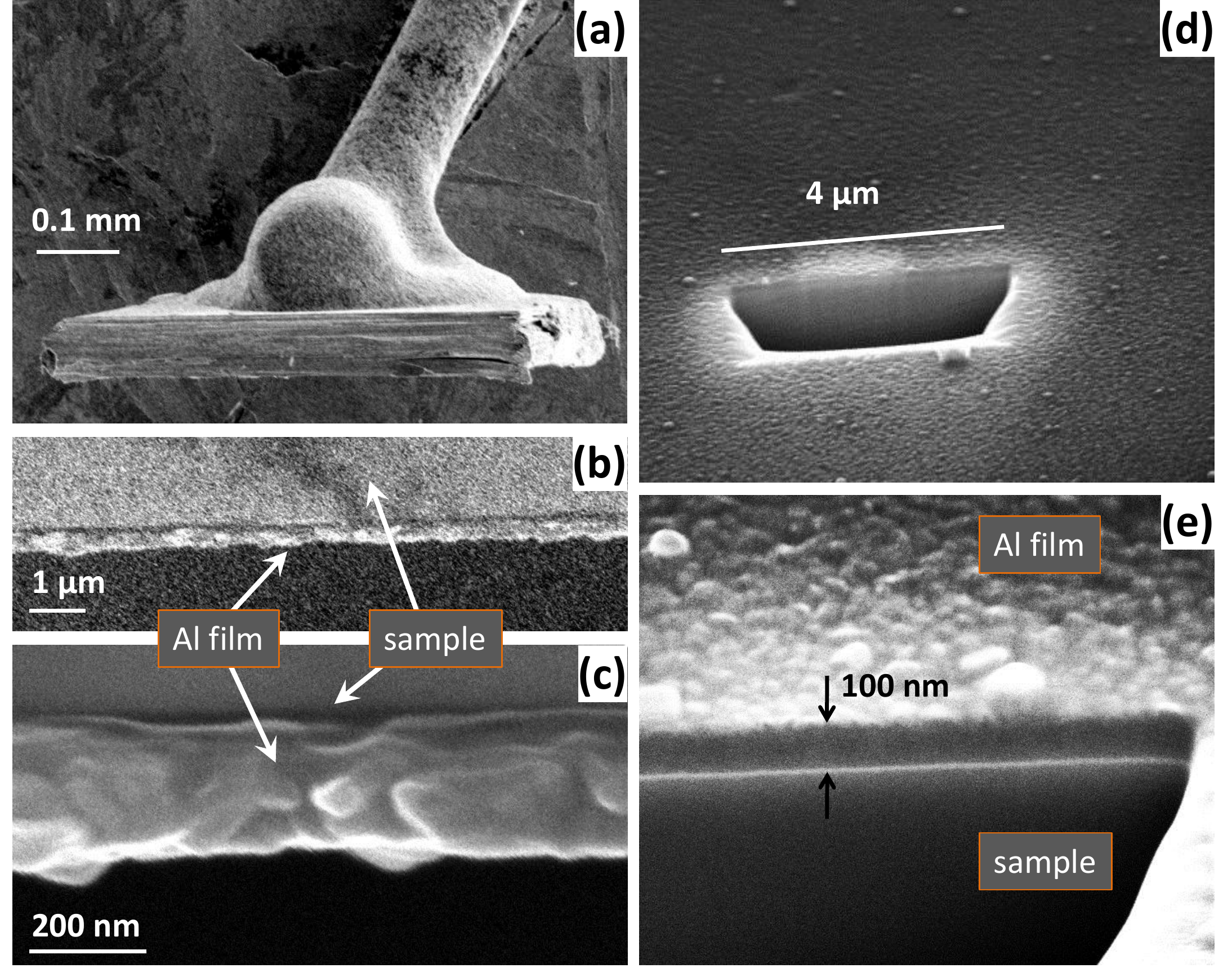}%
\caption{Scanning electron microscope images of the Al coated samples. (a) Large scale view. The broken side is on top. (b) and (c) are zoomed in on the Al film on the edge of the broken side. (d) A trench produced by a focused-ion beam (FIB). (e) Close-up view of the FIB trench showing the Al film and its thickness.}
\label{FeCoFig5}
\end{figure}

To further investigate the effects of doping and the difference between the batches, we use the method described in section \ref{SEC:L0}, which involves measuring the sample, coating it with a uniform layer of Al and re-measuring \cite{Prozorov2000L0,Gordon2010FeCoL0vsX}. The Al film was deposited onto each sample while it was suspended from a rotating stage by a fine wire in an argon atmosphere of a magnetron sputtering system. Film thickness was checked using a scanning electron microscope in two ways, both of which are shown in Fig.~\ref{FeCoFig5}. The first method involved breaking a coated sample after all measurements had been performed to expose its cross section.  After this, it was mounted on an SEM sample holder using silver paste, shown in Fig.~\ref{FeCoFig5}(a). The images of the broken edge are shown for two different zoom levels in Fig.~\ref{FeCoFig5}(b) and (c). The second method used a focused ion beam (FIB) to make a trench on the surface of a coated sample, with the trench depth being much greater than the Al coating thickness, shown in Fig.~\ref{FeCoFig5}(d). The sample was then tilted and imaged by the SEM that is built into the FIB system, shown in Fig.~\ref{FeCoFig5}(e).

\begin{figure}[h]
\centering
\includegraphics[width=9cm]{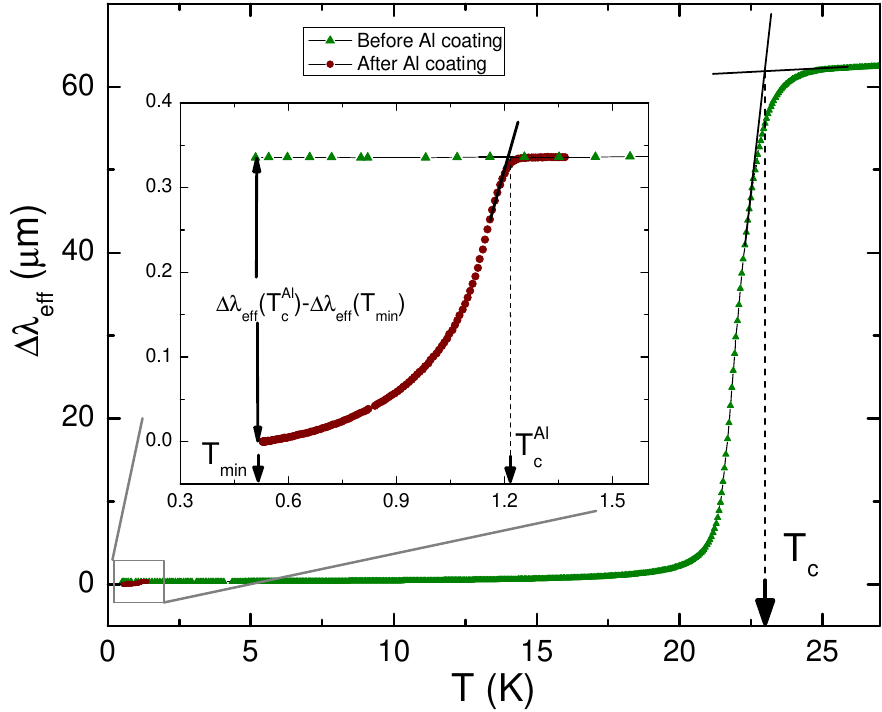}%
\caption{\underline{Main frame}: Full superconducting transition of an optimally doped FeCo122 crystal from batch \#1 before and after Al coating. \underline{Inset}: Zoomed in low-temperature region, $T_{min} \lesssim T \lesssim T_c^{Al}$, before (green triangles) and after (brown circles) the Al coating on the same sample. (Notice how small is the effect on a full scale of the main frame).}
\label{FeCoFig6}
\end{figure}

Example of the penetration depth measurements before and after coating are shown in Fig.~\ref{FeCoFig6}. Notice how small is the effect of Al coating when presented on a large scale of a full superconducting transition of the coated sample. However, TDR technique is well suited to resolve the variation due to aluminum layer \cite{Prozorov2000L0,Gordon2010FeCoL0vsX}.

\begin{figure}[h]
\centering
\includegraphics[width=9cm]{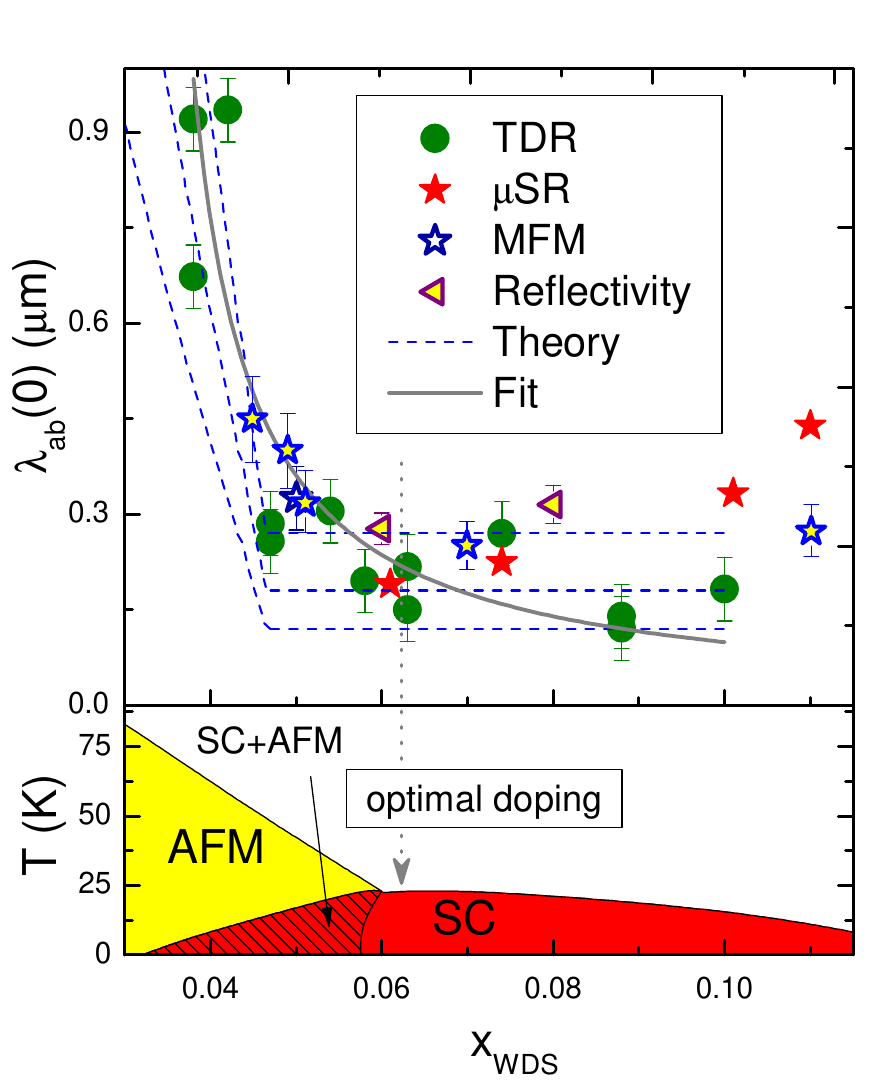}%
\caption{\underline{Top panel}: The zero temperature London penetration depth, $\lambda_{ab}(0)$, as a function of the Co concentration, $x$. The three dashed blue lines are theoretical curves obtained using Eq.~(\ref{SC+SDW}) for three different values of $\lambda_{ab}(0)$ in the pure superconducting state. The solid gray line is a fit to the TDR data only of the form A+B/$x^n$.  Also shown are values of $\lambda_{ab}(0)$ obtained by other experiments for comparison explained in the text. \underline{Bottom panel}: Schematic phase diagram for FeT122 system showing the coexisting region \cite{Fernandes2010unconvPairing,Nandi2010OrthoDist}.}
\label{FeCoFig7}
\end{figure}

Obtained values of $\lambda_{ab}(0)$ are summarized in the top panel of Fig.~\ref{FeCoFig7} for doping levels, $x$, across the superconducting region of the phase diagram, shown schematically in the bottom panel of Fig.~\ref{FeCoFig7}. The size of the error bars for the $\lambda_{ab}(0)$ points was determined by considering the film thickness to be $t=100 \pm10$ nm and $\lambda_{Al}(0)=50\pm 10$ nm. The scatter in the $\lambda_{ab}(0)$ values shown in the upper panel of Fig.~\ref{FeCoFig7} has an approximately constant value of $\pm$ 0.075 $\mu$m for all values of $x$, which probably indicates that the source of the scatter is the same for all samples.  For comparison, Fig.~\ref{FeCoFig7} also shows $\lambda_{ab}(0)$ obtained from $\mu$SR measurements (red stars) \cite{Williams2010FeCouSR}, the MFM technique (open stars) \cite{Luan2010FeCo122Opt,Luan2011FeCoL0}
and optical reflectivity (purple open triangles) \cite{Nakajima2010opticsL0}. Given statistical uncertainty these measurements are consistent with our results within the scatter.  It may also be important to note that the $\lambda_{ab}(0)$ values from other experiments are all on the higher side of the scatter that exists within the TDR $\lambda_{ab}(0)$ data set. As discussed above, our TDR techniques give a low bound of $\lambda(0)$, consistent with this observation.

In order to provide a more quantitative explanation for the observed increase in $\lambda_{ab}(0)$ as $x$ decreases in the underdoped region, we have considered the case of $s^{\pm}$ superconductivity coexisting with itinerant antiferromagnetism \cite{Fernandes2010unconvPairing}. For the case of particle hole symmetry (nested bands), the zero temperature value of the in-plane penetration depth in the region where the two phases coexist is given by \cite{Fernandes2010unconvPairing,Gordon2010FeCoL0vsX}:

\begin{equation}
\label{SC+SDW}
\lambda^{SC+SDW}_{ab}(0)=\lambda^0_{ab}(0)\sqrt{1+\frac{\Delta^2_{AF}}{\Delta^2_0}}
\end{equation}

\noindent where $\lambda^0_{ab}(0)$ is the value for a pure superconducting system with no magnetism present, and $\Delta_{AF}$ and $\Delta_0$ are the zero temperature values of the antiferromagnetic and superconducting gaps, respectively.  Deviations from particle hole symmetry lead to a smaller increase in $\lambda^{SC+SDW}_{ab}(0)$, making the result in Eq.~(\ref{SC+SDW}) an upper estimate \cite{Fernandes2010unconvPairing}.

The three blue dashed lines shown in the top panel of Fig.~\ref{FeCoFig7}, which were produced using Eq.~(\ref{SC+SDW}), show the expected increase in $\lambda_{ab}(0)$ in the region of coexisting phases below $x\approx0.047$ by normalizing to three different values of $\lambda_{ab}(0)$ in the pure superconducting state, with those being 120 nm, 180 nm and 270 nm to account for the quite large dispersion of the experimental values.  This theory does not take into account changes in the pure superconducting state, so for $x>0.047$ the dashed blue lines are horizontal.  These theoretical curves were produced using parameters that agree with the phase diagram in the bottom panel of Fig.~\ref{FeCoFig7} \cite{Fernandes2010unconvPairing,Nandi2010OrthoDist}. While the exact functional form was not provided by any physical motivation and merely serves as a guide to the eye, the solid gray line in Fig.~\ref{FeCoFig7} is a fit of the TDR $\lambda_{ab}(0)$ data to a function of the form A+B/$x^n$, which does indeed show a dramatic increase of $\lambda_{ab}(0)$ in the coexistence region and also a relatively slight change in the pure superconducting phase. It should be noted that a dramatic increase in $\lambda_{ab}(0)$ below $x\approx0.047$ cannot be explained by the impurity scattering, which would only lead to relatively small corrections in $\lambda(T)$.

\begin{figure}[h]
\centering
\includegraphics[width=9cm]{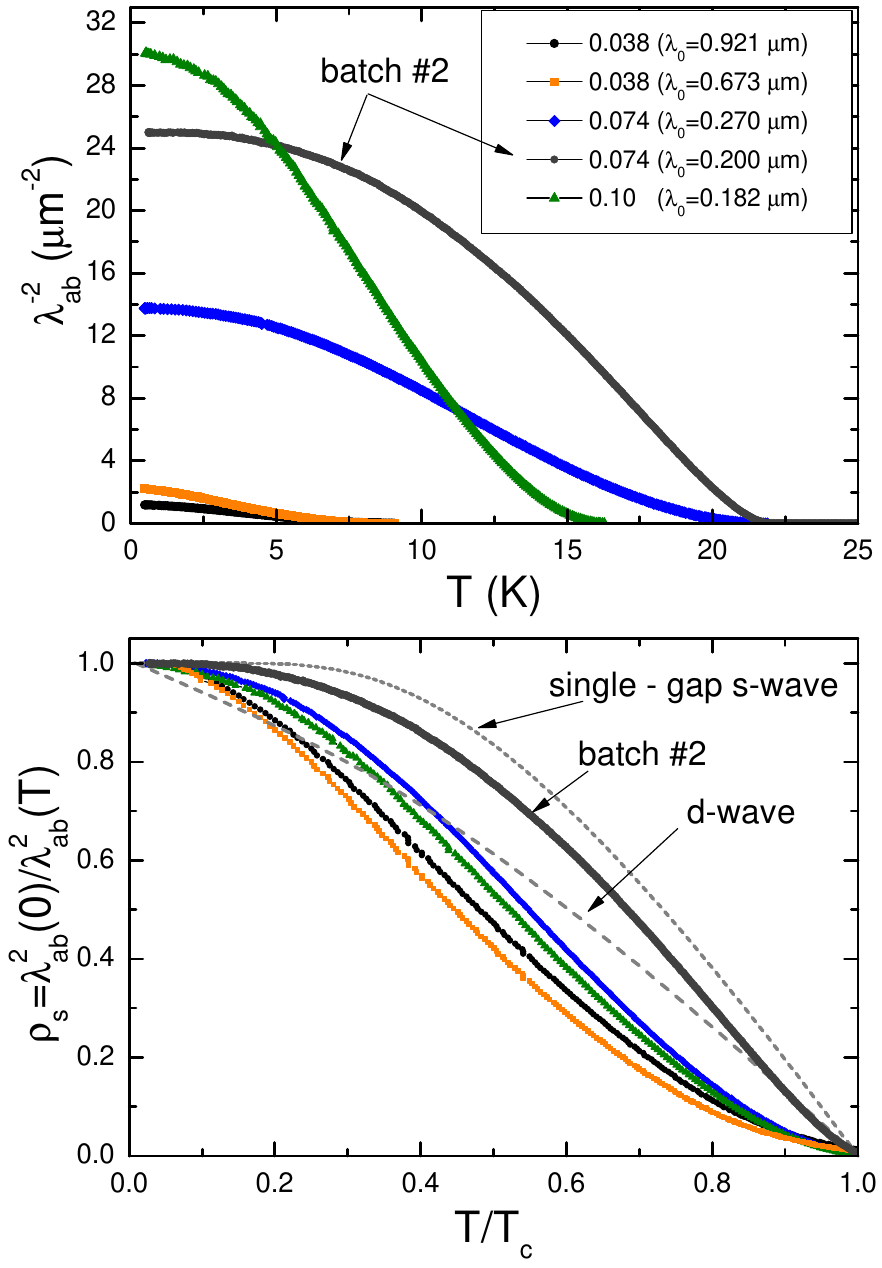}%
\caption{\underline{Top panel}: $\lambda^{-2} \propto n_s$ vs. $T$ for underdoped, optimally doped and overdoped concentrations for batch \#1. A sample from batch \#2 is shown for comparison. \underline{Bottom panel}: normalized superfluid density, $\rho_s \equiv n_s(T)/n_s(0) = (\lambda(0)/\lambda(T))^2$.}
\label{FeCoFig8}
\end{figure}

With the experimental values of $\lambda(0)$ we can now analyze the superfluid density. In general $\lambda^{-2}(T)$ is given by Eq.~(\ref{lambda-tensor}) and depends on the averaging over the particular Fermi surface. For example, in the simplest cylindrical case, Eq.~(\ref{lambda-d}) and $\lambda^{-2}(0)=4 \pi e^2 N(0)v^2/c^2$, where $N(0)$ is the density of states at the Fermi level and $v$ is Fermi velocity. However, it is instructive to analyze the behavior from a two-fluid London theory point of view looking at the density of superconducting electrons, $n_s= (mc^2/4\pi e^2)\lambda^{-2}$ as function of temperature. The zero value, $n_s(0)$ will in general be less than total density of electrons due to pair-breaking scattering, so the magnitude of $n_s$ is useful when comparing different samples.

Figure \ref{FeCoFig8} shows the data for underdoped, optimally doped and overdoped samples from batch \#1 and also a sample from batch \#2 for comparison. For this sample \#2 $\lambda(0) = 200$ nm was used. There is a clear, but expected, asymmetry with respect to doping. Underdoped samples show quite low density, because not all electrons are participating in forming the Cooper pairs and parts of the Fermi surface are gapped by the SDW as was discussed above. The overdoped sample, $x=0.01$, despite having smaller $T_c$ than the ones with $x = 0.074$, shows the highest $n_s$. Obviously, the data scatter is significant. Therefore, the only reliable conclusion is that penetration depth increases dramatically upon entering the coexisting region. The overdoped side has to be studied more to acquire enough statistics. Furthermore, comparing two samples with $x=0.074$ from two different batches reveals an even more striking difference. Not only does the sample from batch \#2 has larger $n$, but the temperature dependence of $\rho_s$ in the full temperature range is also quite different. The pronounces convex shape (positive curvature) of $\rho_s(T)$ observed in all samples from batch \#1 at the elevated temperatures becomes much less visible in sample \#2. The bottom panel of \ref{FeCoFig8} clearly demonstrates this difference, which is hard to understand based purely on the geometrical consideration (different thicknesses). It seems that thicker samples of the batch \#1 had higher chance of being chemically inhomogeneous across the layers. On the other hand, the convex shape of $\rho_s(T)$ at elevated temperatures is a sign of the two-gap superconductivity \cite{Kogan2009gamma}, which depend sensitively on the interaction matrix, $\lambda_{\nu\mu}$, see section \ref{SEC:gamma-model} and Eq.~(\ref{self-cons1}). This feature becomes more pronounced when the interband coupling, $\lambda_{12}$, becomes smaller compared to the in-band coupling potentials, $\lambda_{11}$ and $\lambda_{22}$. If our interpretation that the difference between $x=0.074$ samples from batch \#1 and batch \#2 is due to enhanced pair-breaking scattering in \#1, this would indicate that this \emph{scattering} is primarily of inter-band character, so it disrupts the \emph{inter-band pairing}. This gives indirect leverage to the $s_{\pm}$ scenario where inter-band coupling plays the major role.

\begin{figure}[h]
\centering
\includegraphics[width=9cm]{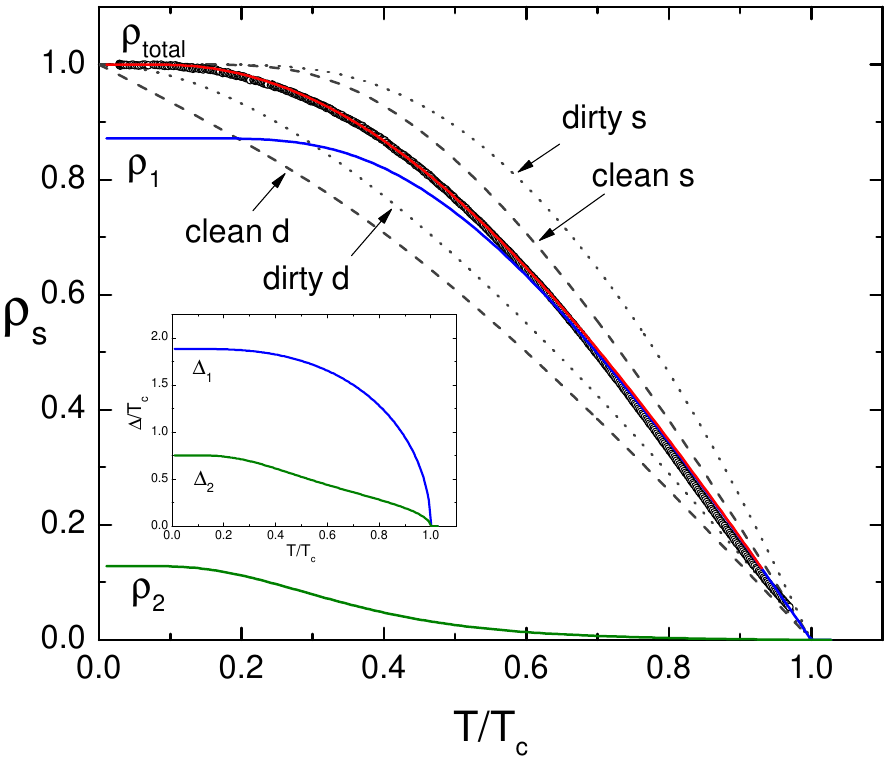}%
\caption{Normalized superfluid density for sample \#2. Symbols show the data and the solid lines represent partial and total $\rho(T)$ obtained from a fit to Eq.(\ref{rhogamma}). Also shown are the clean (dashed grey lines) and dirty (dotted grey lines) single gap $s-$ and $d-$wave cases. Inset shows superconducting gaps obtained self-consistently during the fitting, see Eq.(\ref{.pdf})}.
\label{FeCoFig9}
\end{figure}

The normalized superfluid density for sample \#2 is analyzed in Fig.~\ref{FeCoFig9} by using a two-band $\gamma-$model described is section \ref{SEC:gamma-model}. Symbols show the data and the solid lines represent partial, $\rho_1(T)$ and $\rho_2(T)$, as well as total $\rho_s(T)$ obtained from the fit using Eq.(\ref{rhogamma}). The fit requires solving the self-consistent coupled ``gap'' equations, Eq.~(\ref{.pdf}), which are shown in the inset in Fig.~\ref{FeCoFig9}. Parameters of the fit are as follows: $\lambda_11 = 0.80$, $\lambda_22 = 0.49$, $\lambda_12 = 0.061$, $\gamma = 0.87$. We used a Debye temperature of 250 K \cite{Johnston2010review} to obtain the correct $T_c$ via Eq.~\ref{Tc}) that fixed $\lambda_{11}$ and gave $\tilde\lambda = 0.41$. We also assumed equal partial densities of states on the two bands, so the value of $\gamma = 0.87$ most likely comes from the difference in the $k-$ dependent Fermi velocities, but may also reflect the fact that densities of states are not equal. Indeed, presented fitting parameters should not be taken too literally. The superfluid density is calculated from the temperature - dependent gaps (inset in Fig.~\ref{FeCoFig9}) and close temperature dependencies can be obtained with quite different fitting parameters. However, the gaps fully determine the experimental $\rho_s$ and this is the main result. We find that $\Delta_1(0) = 1.883 T_c = 3.73$ meV and $\Delta_2(0) = 0.754 T_c = 1.49$ meV, which are in good agreement with specific heat \cite{Hardy2010CpFeCo,Gofryk2010Cp} and $\mu$SR penetration depth measurements \cite{Williams2009uSRL0} done on the samples of similar composition. Also shown in Fig.~\ref{FeCoFig9}) are the clean (dashed grey lines) and dirty (dotted grey lines) single gap $s-$ and $d-$wave cases. (Note that while the gap does not depend on the non-magnetic impurities in isotropic $s-$wave case (Anderson theorem), the superfluid density does.) Clearly, $\rho_s(T)$ for sample \#2 (and, of course for samples of batch \#1) does not come even close to any of these single-gap scenarios.

\begin{figure}[h]
\centering
\includegraphics[width=9cm]{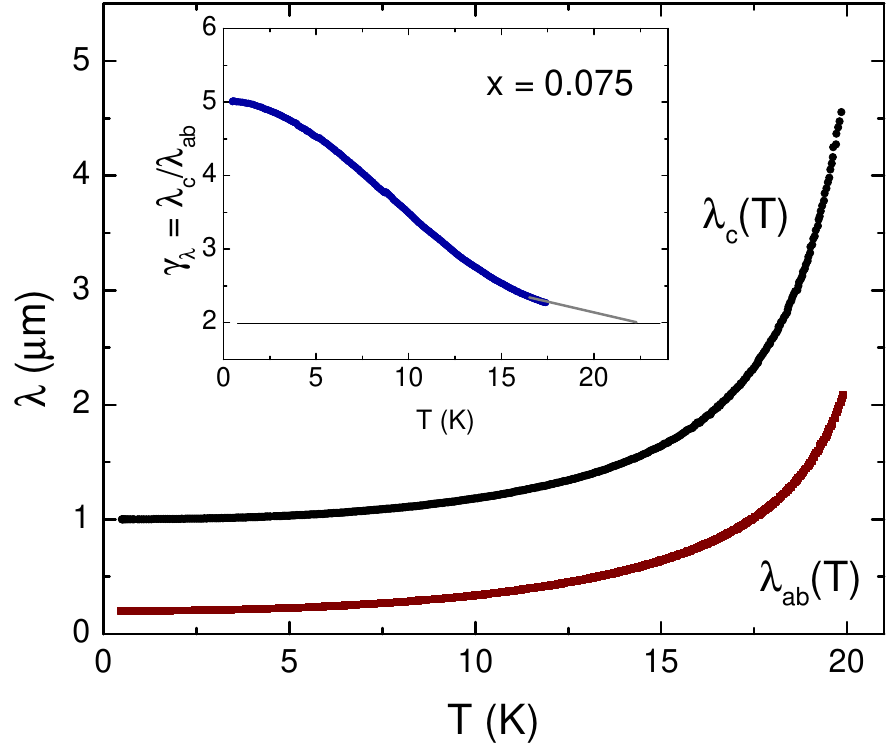}%
\caption{In - plane, $\lambda_{ab}(T)$, and out - of - plane, $\lambda_{c}(T)$ in a single crystal of FeCo122 with $x=0.075$. Inset - temperature - dependent anisotropy, $\gamma_{\lambda} = \lambda_{c}/\lambda_{ab}$.}
\label{FeCoFig10}
\end{figure}

\subsection{Anisotropy of London penetration depths}

\paragraph{}
Let us now discuss the electromagnetic anisotropy in the superconducting state, parameterized by the ratio $\gamma_{\lambda} = \lambda_{c}/\lambda_{ab}$. To determine $\lambda_{c}$ we used method described in section \ref{SEC:Lc} and the results are presented in Fig.~\ref{FeCoFig10}. The problem is that we do not know the absolute value of $\lambda_{c}(0)$, so we could only obtain $\Delta \lambda_{c}(T)$ with the help of knowing $\Delta \lambda_{ab}(T)$, which was measured on the same crystal independently. To find the total $\lambda_{c}(T)$ we use the fact that close to $T_c$ (in the region of validity of Ginzburg-Landau theory),
we should have \cite{Kogan2002anisotropy}:

\begin{equation}
\gamma_\lambda \left( T_c \right) = \sqrt{\gamma_{\rho}\left(T_c \right)}
\end{equation}

\noindent where anisotropy of normal state resistivity, $\gamma_{\rho} = \rho_c/\rho_{ab}$ is taken right above $T_c$. With $\gamma \rho \approx 4 \pm 1$ \cite{Tanatar2009anisotropyRHO}, so that $\gamma_\lambda \left( {T_c} \right) \approx 2$. The results is show in the inset in Fig.~\ref{FeCoFig10}. Of course, there is some ambiguity in determining the exact anisotropy value, but the qualitative behavior does not change - the anisotropy \emph{increases} upon cooling. With our estimate it reaches a modest value of 5 at low temperatures, which makes pnictides very different from high-$T_c$ cuprates. This is opposite to a two-gap superconductor MgB$_2$ where $\gamma_{\lambda}$ \emph{decreases} upon cooling \cite{Kogan2004MgB2,Fletcher2005MgB2}, which may be due to different dimensionality of the Fermi sheets. In any case, temperature - dependent $\gamma_{\lambda}$ can only arise in the case of a multi-gap superconductor.

\begin{figure}[h]
\centering
\includegraphics[width=9cm]{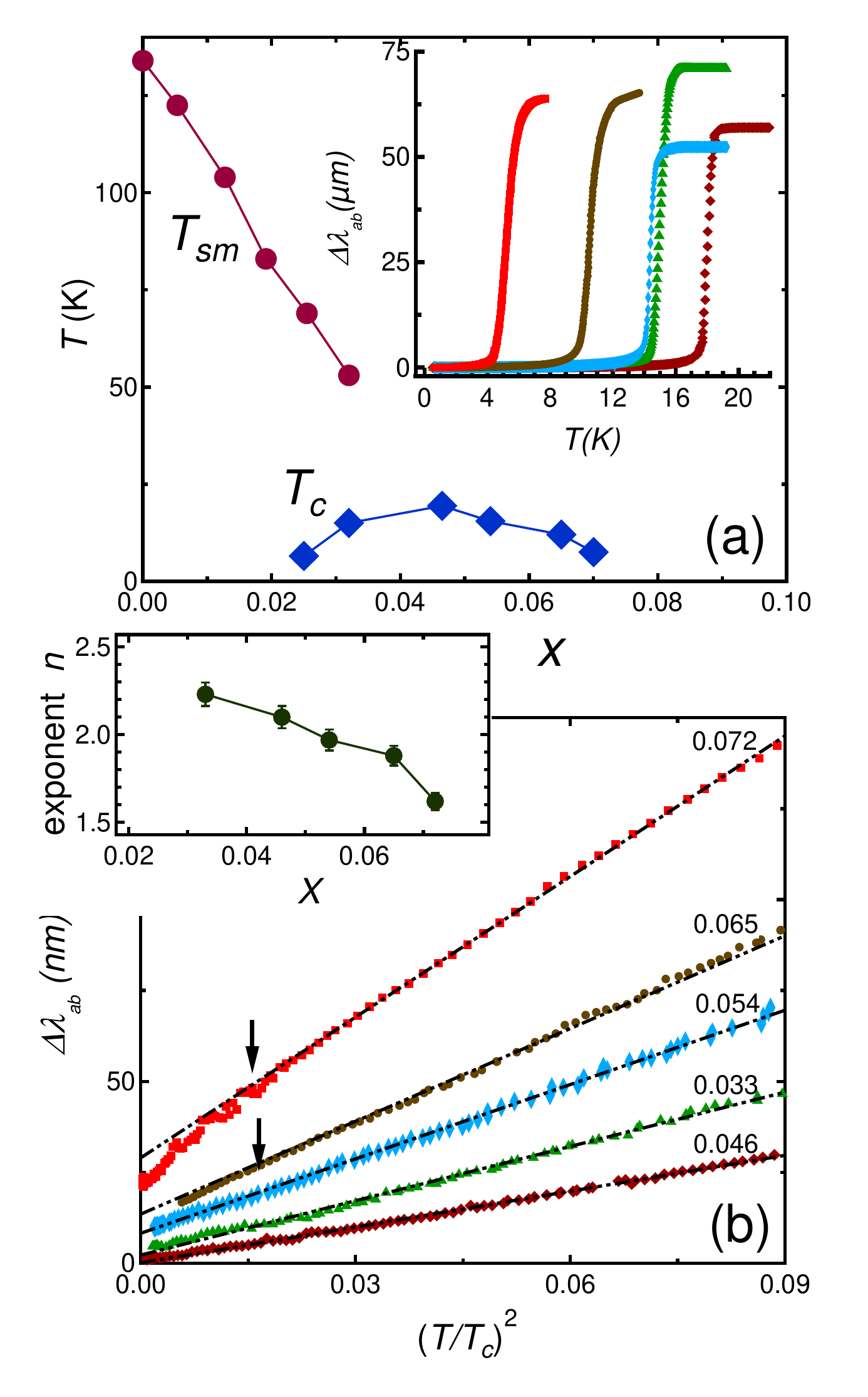}%
\caption{(a) Temperature-doping phase diagram of Ba(Fe$_{1-x}$Ni$_x$)$_2$As$_2$. The inset shows full-temperature range superconducting transitions. (b) $\Delta \lambda _{ab} (T)$ for different doping levels vs. $(T/T_{c})^{2}$. The arrows mark the temperature below which $n$ becomes less than 2. The curves are shifted vertically for clarity. The inset shows the power-law exponent $n(x)$ obtained by fitting to $\lambda _{ab} (T)= a+bT^n$ from the base temperature up to $T/T_c$=0.3.}
\label{FeNi122c1}
\end{figure}

Next we examine the anisotropy of $\lambda(T)$ at different doping levels. This study was performed on FeNi122 samples and is reported in detail elsewhere \cite{Martin2010FeNinodes}. Figure~\ref{FeNi122c1}(a) summarizes the $T(x)$ phase diagram showing structural/magnetic ($T_{sm}$) and superconducting ($T_{c}$) transitions. The inset shows TDR measurements in a full temperature range for all concentrations used in this study. Figure~\ref{FeNi122c1}(b) shows the low-temperature ($T\leq 0.3T_{c}$) behavior of $\lambda_{ab}(T)$ for several Ni concentrations. The data plotted versus $(T/T_c)^2$ are linear for underdoped compositions and show a clear deviation towards a smaller power-law exponent (below temperatures marked by arrows in Fig.~\ref{FeNi122c1}(b)) for overdoped samples. While at moderate doping levels the results are fully consistent with our previous measurements in FeCo-122 \cite{Gordon2009FeCo122vsX,Gordon2009FeCo122Opt}, the behavior in the overdoped samples is clearly less quadratic. It should be noticed that in order to suppress $T_c$ by the same amount, one needs a two times lower Ni concentration compared to Co. In previously FeCo-122 \cite{Gordon2009FeCo122vsX}, the samples never reached highly overdoped compositions equivalent to $x=0.072$ of Ni shown in Fig.~\ref{FeNi122c1}(b). Therefore, Ni doping has the advantage of spanning the phase diagram with smaller concentrations of dopant ions, which may act as the scattering centers. The evolution of the exponent $n$ with $x$ is summarized in the lower inset in Fig.~\ref{FeNi122c1}.

\begin{figure}[h]
\centering
\includegraphics[width=9cm]{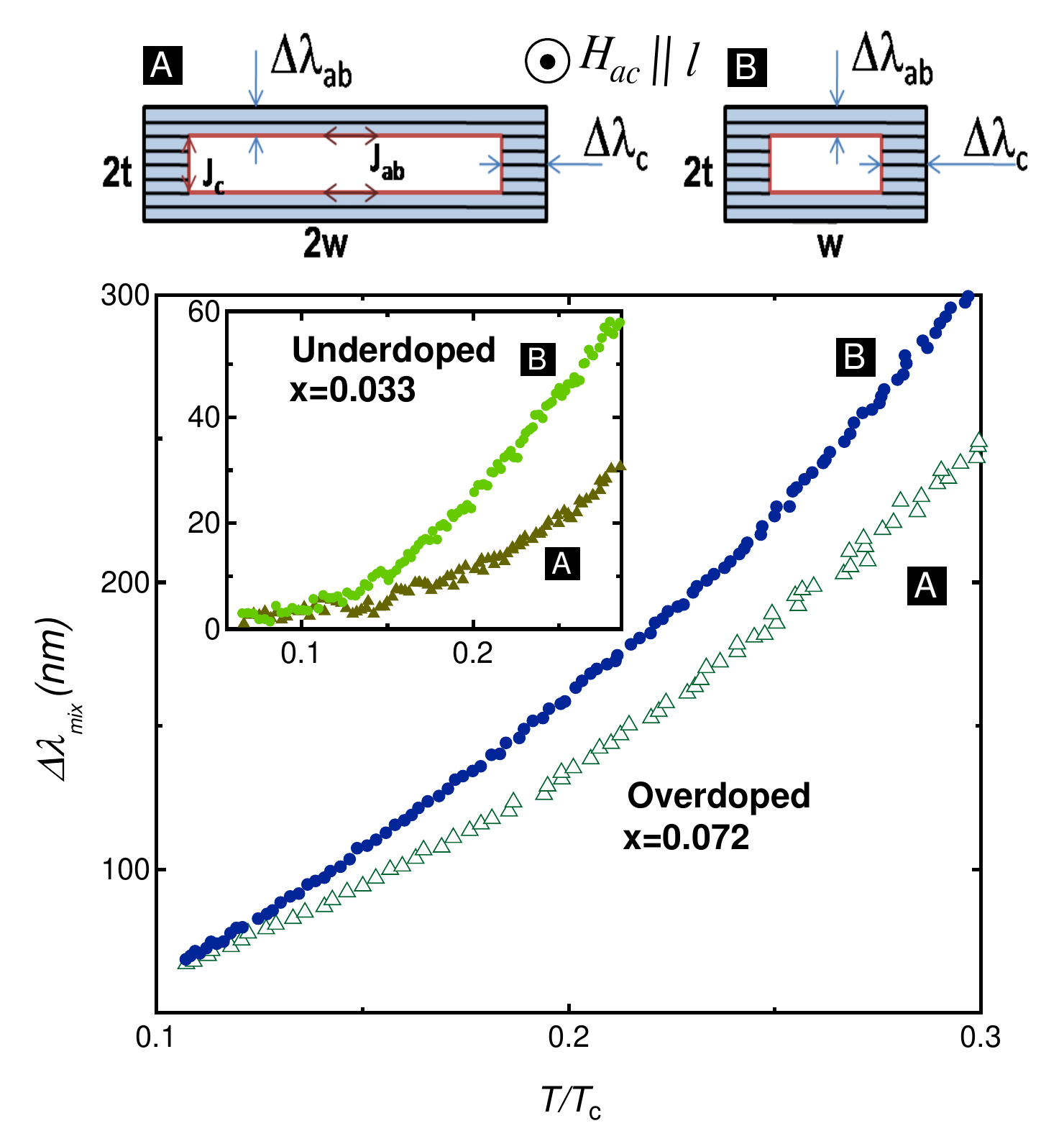}%
\caption{Schematics of magnetic field penetration in the case of $H_{ac} \parallel l$ for the whole sample [A] and after cutting in half along length $l$ [B]. (Main panel) The mixed penetration depth $\Delta\lambda_{mix}(T)$ before [A] and after [B] cutting for the overdoped sample $x$=0.072, $T_c=7.5$ K. Inset shows similar data for the underdoped $x$=0.033, $T_c=15$ K.}
\label{FeNi122c3}
\end{figure}

\begin{figure}[h]
\centering
\includegraphics[width=9cm]{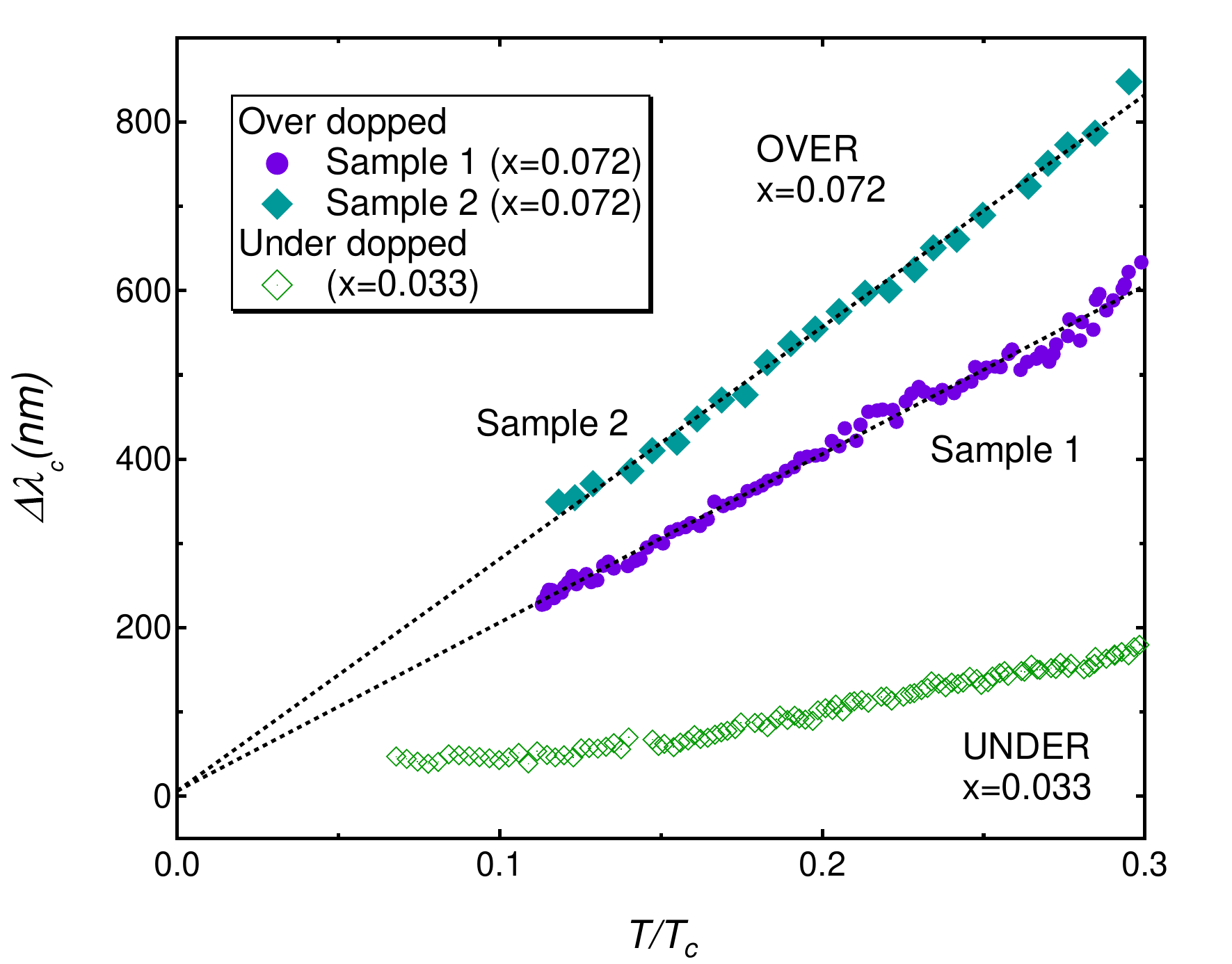}%
\caption{$\Delta\lambda _{c}(T)$ for the underdoped, $x=0.033$, $T_c=15$ K (UNDER), and for two overdoped, $x=0.072$, $T_c=7.5$ K and $T_c=6.5$ K (OVER), samples. Dashed lines are linear fits.}
\label{FeNi122c4}
\end{figure}

Now we apply a technique described in Sec.~\ref{SEC:Lc} to estimate $\lambda_{c}(T)$. Figure~\ref{FeNi122c3} shows the effective penetration depth, $\lambda_{mix}$ (see Eq.~\ref{eqmix}), for overdoped, $x$=0.072 (main panel), and underdoped, $x$=0.033 (inset), samples before (A) and after (B) cutting in half along the longest side ($l$-side) as illustrated schematically at the top of the figure. Already in the raw data, it is apparent that the overdoped sample exhibits a much smaller exponent $n$ compared to the $\lambda_{ab}(T)$, while underdoped samples show a tendency to saturate below 0.13T$_{c}$. Using Eq.~\ref{eqmix} we can now extract the true temperature dependent $\Delta \lambda_c(T)$. The result is shown in Fig.~\ref{FeNi122c4} for two different overdoped samples of the same composition, $x=0.072$ having $T_c=7.5$ K and $T_c=6.5$ K, and for an underdoped sample with $x=0.033$ having $T_c=15$ K. Since the thickness of the sample is smaller than its width, we estimate the resolution of this procedure for $\Delta\lambda _{c}$ to be about 10 nm, which is much lower than $0.2$ nm for $\Delta\lambda_{ab}$. Nevertheless, the difference between the samples is obvious. The overdoped samples show a clear linear temperature variation up to $T_c/3$, strongly suggesting nodes in the superconducting gap.  The average slope is large, about $d\lambda_{c}/dT\approx 300$ nm/K indicating a significant amount of thermally excited quasiparticles. By contrast, in the underdoped sample the inter-plane penetration depth saturates indicating a fully gapped state. If fitted to the power-law the exponent in the underdoped sample $2\leq n\leq 3$, depending on the fitting range.

Nodes, if present somewhere on the Fermi surface, affect the temperature dependence of both components of $\Delta\lambda(T)$. However, the major contribution still comes from the direction of the supercurrent flow, thus placing the nodes in the present case at or close to the poles of the Fermi surface. The nodal topologies that are consistent with our experimental results are latitudinal circular line nodes located at the finite $k_z$ wave vector or a point (or extended area) polar node with a nonlinear ($\Delta(\theta) \sim \theta^p$, $p>1$) variation of the superconducting gap with the polar angle, $\theta$. It is interesting to note a close similarity to the results of thermal conductivity measurements in overdoped FeCo122 that have reached the same conclusions - in - plane state is anisotropic, but nodeless \cite{Tanatar2010FeCoTCvsX}, whereas out of plane response is nodal \cite{Reid2010FeCo122TC}. Still, we emphasize that the apparent power-law behavior of the in-plane penetration depth, $\lambda_{ab}(T)$, in a heavily overdoped samples could be induced by the out-of-plane nodes \cite{Graser2010,Mishra2011}. To summarize, it appears that not only is the gap not universal across different pnictide families \cite{Hashimoto2010AsPnodesPD}, it is not universal even within the same family over an extended doping range. Similar conclusion has been reached for the hole-doped BaK122 pnictides \cite{Thomale2011,Kim2011BaKLambda,Reid2011BaK122lineNodes}.

\subsection{Pair-breaking}
\paragraph{}

Although the natural variation in the scattering rates between samples provides a good hint towards importance of pair-breaking scattering, for more quantitative conclusions we need to introduce additional disorder. This is can be achieved with the help of heavy-ion irradiation. To examine the effect of irradiation, $\sim 2\times 0.5\times 0.02-0.05$ mm$^3$ single crystals were
selected and then cut into several pieces preserving the width and the thickness. We compare sets of
samples, where the samples in each set are parts of the same original large crystal and had identical temperature-dependent penetration depth in unirradiated state. (These samples is what we call batch \#2 in this review with unirradiated reference piece appearing in the discussion and figures of the previous sections). Irradiation with 1.4 GeV $^{208}$Pb$^{56+}$ ions was performed at the Argonne Tandem Linear Accelerator System (ATLAS) with an
ion flux of $\sim 5\times 10^{11}$ ions$\cdot$s$^{-1}$$\cdot$m$^{-2}$. The actual total dose was
recorded in each run. Such irradiation usually produces columnar defects or the elongated pockets of disturbed superconductivity along the ions propagation direction. The density of defects, $d$, per unit area is usually expressed in terms of so-called ``matching field'', $B_\phi=\Phi_0 d$, which is obtained assuming one flux quanta,
$\Phi_0\approx 2.07\times 10^{-7}$ G$\cdot$cm$^2$ per ion track. Here we studied samples with
$B_\phi=0.5$, 1.0 and 2.0 T corresponding to $d=2.4\times 10^{10}$ cm$^{-2}$, $4.8\times 10^{10}$
cm$^{-2}$ and $9.7\times 10^{10}$ cm$^{-2}$. The sample thickness was chosen in the range of
$\sim 20 -50 \mu$m to be smaller than the ion penetration depth, $60 - 70~\mu$m. The same samples were studied by
magneto-optical imaging.  The strong Meissner screening and large uniform enhancement of pinning have
shown that the irradiation has produced uniformly distributed defects \cite{Prozorov2010irr}.

\begin{figure}[h]
\centering
\includegraphics[width=9cm]{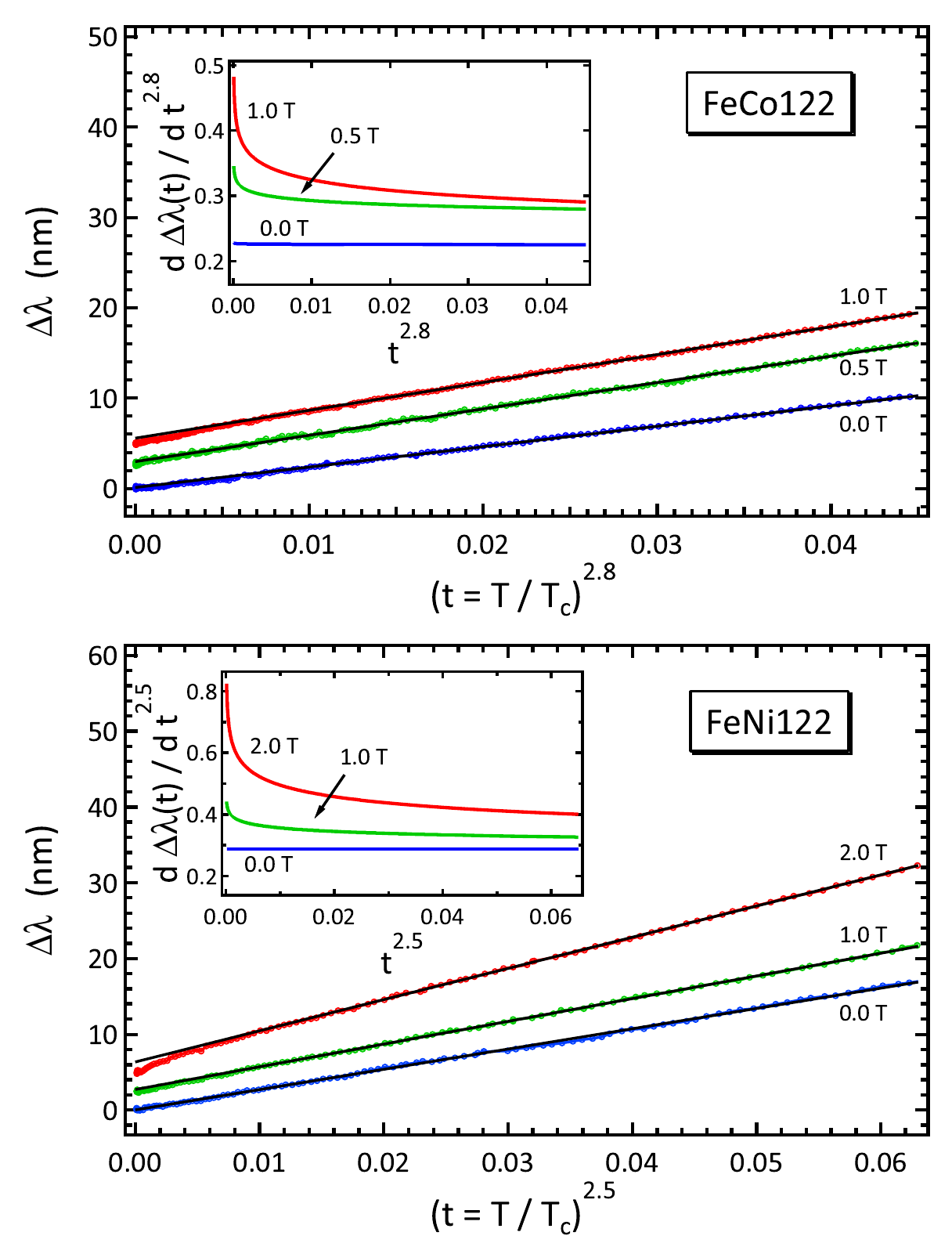}
\caption{Detailed comparison of the functional form of $\Delta\lambda(T)$ for irradiated FeCo122
and FeNi122. In the main panels $\Delta\lambda(T)$ is plotted vs. $(t=T/T_c)^{n_0}$ with the
exponents $n$ taken from the best fits of \emph{unirradiated} samples: $n_0 =$ 2.8 and 2.5 for
FeCo122 and FeNi122, respectively. Apparently, irradiation causes
low-temperature deviations, which are better seen in the derivatives, $d \Delta\lambda(t)/dt^{n_0}$,
plotted in the insets.}
\label{FeCo122irr2}
\end{figure}

\begin{figure}[h]
\centering
\includegraphics[width=9cm]{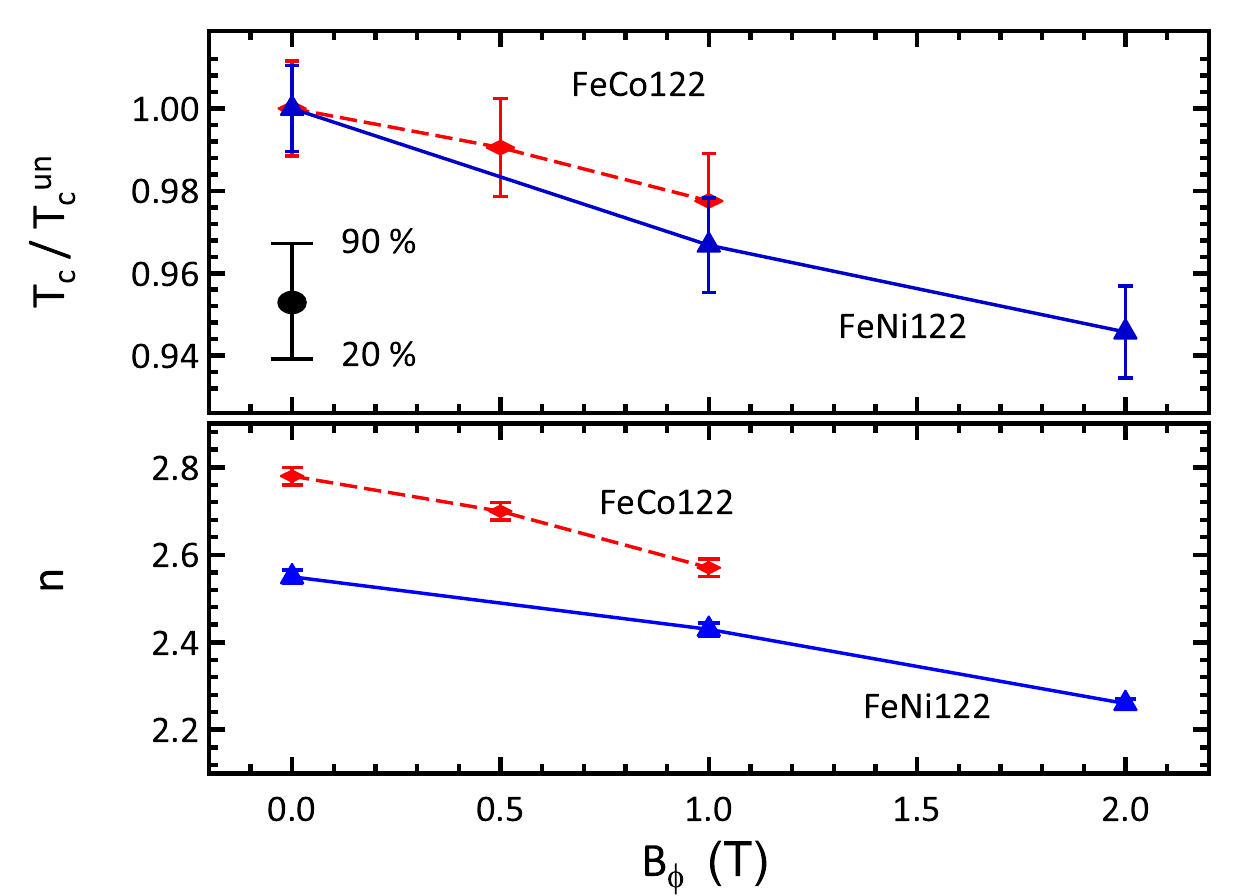}
\caption{Top panel: The suppression of $T_c$ with disorder relative to
\emph{unirradiated} $T^{un}_c$. The vertical bars denote the width of the transition corresponding the diamagnetic signal change
from 90\% (onset) to 20\% (end). Symbols are shown at the mean values between onset and end of the transition. Lower panel: exponent $n$ vs. $B_{\phi}$.}
\label{FeCo122irr3}
\end{figure}

Indeed, to see the effect, we need to start with the best (largest exponent) sample we have. We irradiated the sample designated as batch \#2 with $n=2.83$, which was discussed in detail above (see Figures \ref{FeCoFig3}, \ref{FeCoFig4} and \ref{FeCoFig9}).
To analyze the power-law behavior and its variation with irradiation, we plot $\Delta\lambda$ as
a function of  $(t=T/T_c)^{n_0}$ in Fig.~\ref{FeCo122irr2}, where the $n_0$ values for FeCo122 and FeNi122 were
chosen from the best power-law fits of the unirradiated samples (see Fig.~\ref{FeCo122irr3}). While the data
for unirradiated samples appear as almost perfect straight lines showing robust power-law behavior, the
curves for irradiated samples show downturns at low temperatures indicating smaller exponents. This
observation, emphasized by the plots of the derivatives $d\Delta\lambda(t)/dt^{n_0}$ in the inset of
Fig.~\ref{FeCo122irr2}, points to a significant change in the low-energy excitations with radiation.

The variations of $T_c$ and $n$ upon irradiation are illustrated in
Fig.~\ref{FeCo122irr3}. Dashed lines and circles show FeCo122, while solid lines and
triangles show FeNi122. The upper panel shows the variation of $T_c$ and the width of
the transition. Since $B_{\phi}$ is directly proportional to the area density of the ions, $d$, we can say that $T_c$ decreases roughly linearly with $d$. The same trend is evident for the exponent $n$ shown in the lower panel of Fig.~\ref{FeCo122irr3}.

The influence of impurities, assuming $s_{\pm}$ pairing, has been analyzed numerically in a T-matrix approximation \cite{Kim2010irr}.  Figure \ref{FeCo122irr3}(a) shows calculated superfluid densities for different values of the scattering rate. Figure \ref{FeCo122irr3}(b) shows corresponding densities of states. Finally, Fig.~\ref{FeCo122irr3}(c) shows the central result: the correlation between $T_c$ and $n$. Note that these two quantities are obtained independently of each other. Assuming that the unirradiated samples
have some disorder due to doping, and scaling $T^{un}_c$ to lie on the theoretical curve, we find that the $T_c(B_{\phi})$ of the irradiated samples also follows this curve. The assumption of similarity between doping and radiation-induced disorder, implied in this comparison, while not unreasonable, deserves further scrutiny. More recent discussion on the variation of $T_c$ with disorder is found in Ref.~(\cite{Efremov2011TcVsDisorder}).

\begin{figure}[h]
\centering
\includegraphics[width=9cm]{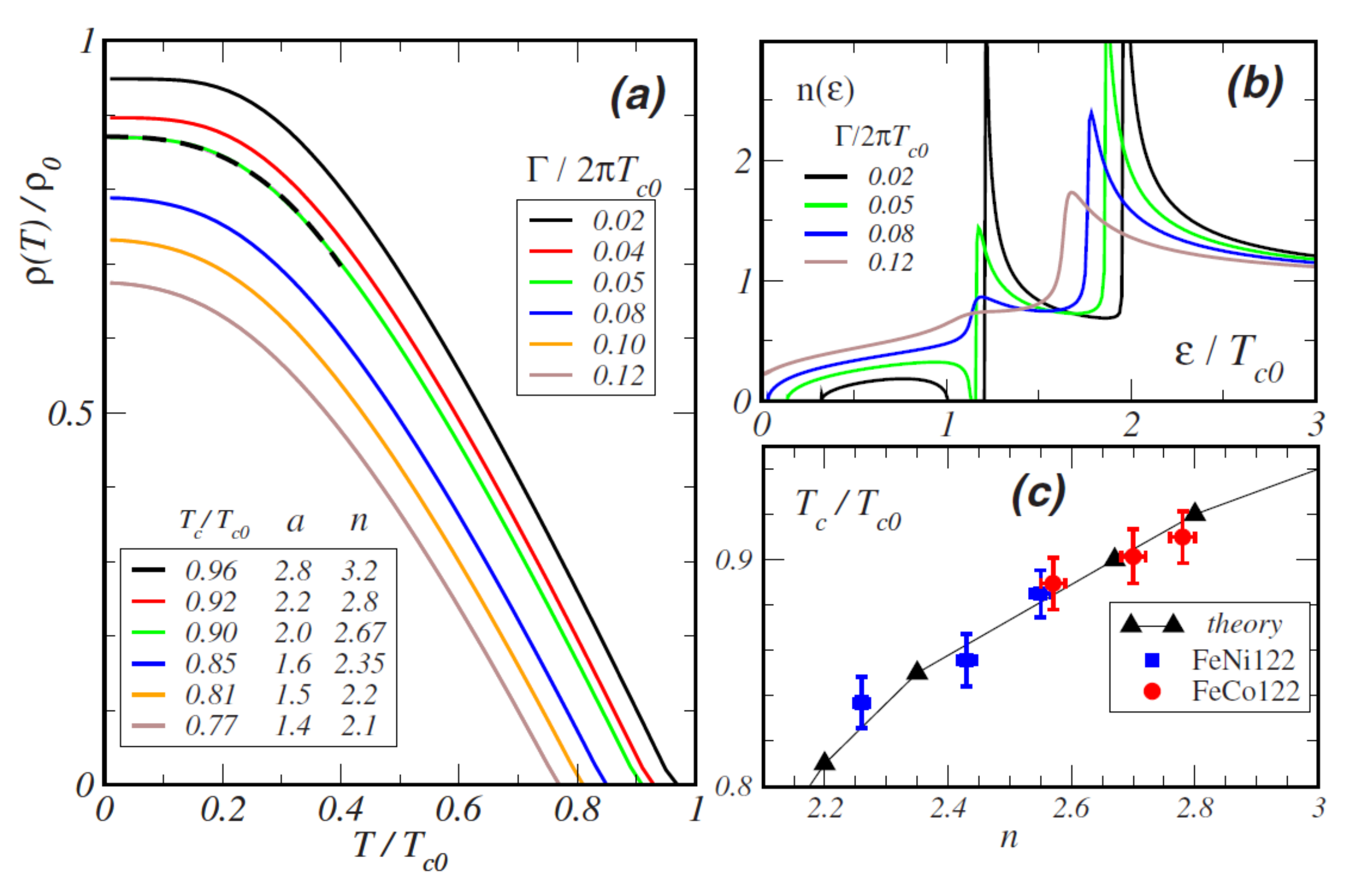}%
\caption{(a) Superfluid density and (b) the density of states,
computed for the $s_{\pm}$ state with sign-changing isotropic gaps
and strong interband impurity scattering, between the Born and unitary limits.
The dashed line in (a) is an example of a power-law fit
$\rho(T)/\rho_0 = \rho(0)/\rho_0 - a (T/T_{c0})^n$ for $0<T <0.4 \, T_{c0}$;
(b) As the impurity concentration $n_{imp} \sim \Gamma$)) increases,
the band of mid-gap states approaches the Fermi level and the exponent $n$ is reduced.
(c) $T_c$ vs. power $n$, from the theoretical model (triangles)
and experiment (squares and circles).}
\label{FeCo122irr4}
\end{figure}

\section{Conclusions}
\paragraph{}
It was not possible to include in this review many interesting results obtained for various members of the diverse family of iron-based superconductors. However, we may provide some general conclusions based on our work as well as on results by others.
\begin{enumerate}
  \item The superconducting gap in optimally doped pnictides is isotropic and nodeless.
  \item Pair-breaking scattering changes the clean - limit low temperature asymptotics (exponential for nodeless and $T-$linear for line nodes) to a $\sim T^2$ behavior. Therefore, additional measurements (such as deliberately introduced disorder) are needed to make conclusions about the order parameter symmetry. In anisotropic superconductors in general and in $s_{\pm}$, in particular, even the non-magnetic impurities are pair-breaking.
  \item The materials can be described within a self-consistent two-band $\gamma-$model with two gaps with the  ratio of magnitudes of about $\Delta_1(0)/\Delta_2(0) \approx 2 - 3$.
  \item Upon doping, the power-law exponent, $n$ for the in-plane penetration depth, $\lambda_ab(T)$, decreases reaching values below 2 signaling of developing significant anisotropy, whereas out-of-plane $\lambda_c(T)$ shows a linear~$-T$ behavior signaling of line nodes with Fermi velocity predominantly in the $c-$direction.
  \item There is a fairly large region of coexisting superconductivity and long-range magnetic order, albeit with suppressed superfluid density.
  \item Overall, the observed behavior is consistent with the $s_{\pm}$ pairing, but realistic 3D calculations are required to achieve agreement with experiments.
\end{enumerate}

\section*{Acknowledgments}
This review is based on the experimental results obtained by the members of R.P. laboratory: Makariy Tanatar, Catalin Martin, Kyuil Cho, Ryan Gordon and Hyunsoo Kim during 2008 - 2010. More details can be found in Ryan Gordon's Ph.D. thesis \cite{RyanThesis2011}. Our colleague, Makariy Tanatar, was responsible for all sample preparation and handling.
The samples were grown in the group of Paul Canfield and Sergey Bud'ko. We are grateful to many colleagues for insightful discussions - too many to be listed in the limited space of this review. This research was supported by the U.S. Department of Energy, Office of Basic Energy Sciences, Division of Materials Sciences and Engineering under contract No. DE-AC02-07CH11358. R.P. acknowledges support from the Alfred P. Sloan Foundation.

\section*{References}
\bibliographystyle{unsrt}
\bibliography{LambdaReviewShort}

\end{document}